# Dual deconvolution multifocal fluorescence response matrix microscopy for aberration-free high-throughput super-resolution imaging

Sumin Lim[1*], Sanghyeon Park[1], Tran Thinh Le[1], Jin Hee Hong[1], Seokchan Yoon[2*], Wonshik Choi[1*]

[1]*Department of Physics, Korea University, Seoul 02855, Korea*

[2]*School of Biomedical Convergence Engineering, Pusan National University, Yangsan 50612, Korea*

*Correspondence: lsm9333@naver.com, sc.yoon@pusan.ac.kr, wonshik@korea.ac.kr

## Abstract

Fluorescence response matrix imaging enables separate computational correction of excitation and emission aberrations through dual deconvolution, but single-focus acquisition requires a large number of camera frames, limiting its practical use. Multifocal excitation can accelerate acquisition by recording many illumination positions in parallel, but denser foci also increase inter-focus crosstalk, which can bias the estimated excitation and emission optical transfer functions and degrade aberration correction. Here, we introduce high-throughput aberration-free fluorescence response matrix imaging with adaptive multifocal illumination, in which the inter-focus spacing of the multifocal lattice is adjusted according to specimen-dependent crosstalk. A programmable digital micromirror device generates multifocal excitation lattices, and the resulting detector-resolved fluorescence responses are assembled into a two-coordinate fluorescence response matrix. Dual deconvolution separately estimates and corrects the excitation and emission optical transfer functions, and the aberration-corrected matrix is then combined in the spatial-frequency domain to reconstruct the aberration-free image. This strategy reduces the number of required camera frames by approximately three orders of magnitude, allowing response matrices with ~$10^6$ illumination positions to be acquired within seconds to a few minutes. The accelerated acquisition preserves the aberration-correction capability of dual deconvolution and achieves an estimated lateral resolution of ~100 nm. Applied to heterogeneous mouse brain and kidney tissue, the method enables high-resolution, depth-resolved volumetric imaging while correcting specimen-dependent aberrations. Together, these results establish adaptive multifocal illumination as a practical route to high-throughput aberration-free fluorescence response matrix imaging.

## Introduction

Fluorescence super-resolution microscopy reveals the structural organization of biological specimens beyond the diffraction limit[1,2,3,4], but in optically heterogeneous samples its performance is limited by aberrations[5]. Aberrations distort the point-spread function (PSF), reduce contrast and suppress the high-spatial-frequency information needed to resolve fine structures. Hardware adaptive optics (AO) can compensate for these distortions using deformable mirrors or spatial light modulators[6,7,8], but requires additional correction optics. Direct wavefront-sensing and guide-star implementations can also be difficult to use in densely labeled specimens[9,10]. Matrix-based computational AO approaches use a response matrix indexed by distinct illumination and detection coordinates to estimate aberration-induced distortions and recover aberration-corrected images, as demonstrated in coherent reflection-matrix imaging[11,12] and incoherent fluorescence imaging[13].

Dual deconvolution[14] estimates excitation and emission aberrations separately from an incoherent fluorescence response matrix. Within a locally shift-invariant region, the Fourier-transformed matrix factorizes into an emission optical transfer function (OTF), the object spectrum and an excitation OTF. Estimating these factors requires Nyquist acquisition of the illumination coordinate together with a detection region large enough to retain the aberration-broadened fluorescence response at each position. Because single-focus acquisition samples only one illumination coordinate per camera exposure, this dense sampling translates directly into a very large frameshift count. For a roughly 100 × 100 μm² field sampled at 90–120-nm illumination-sampling intervals, acquisition of the full fluorescence response matrix would require about $10^6$ frames per plane. Nyquist acquisition has therefore remained a major bottleneck for high-throughput dual deconvolution. Multifocal structured illumination microscopy (MSIM) and related multifocal image-scanning microscopy (MISM) approaches provide a natural route to accelerate this acquisition by exciting sparse arrays of foci and recording the corresponding fluorescence responses in parallel[15,16,17,18,19]. For the same illumination-sampling interval and field of view, the required frame count is ideally reduced in proportion to the number of foci recorded simultaneously. Related approaches include spinning-disk image scanning microscopy (ISM)[20], instant structured illumination microscopy with analog processing[21] and multifocal metalens-based ISM[22].

However, increasing focal density both accelerates acquisition and increases inter-focus crosstalk, creating a trade-off between acquisition throughput and reliable aberration estimation. In conventional multifocal pixel-reassignment reconstruction, such crosstalk typically manifests as spatial-domain degradation, including blurring, loss of contrast, and reconstruction artifacts. In dual deconvolution, the same crosstalk more directly compromises aberration correction. Fluorescence originating from a neighboring focus can be assigned to the wrong illumination position, thereby biasing the separately estimated excitation and emission transfer functions. Because the degree of response crosstalk depends on specimen-dependent aberration, scattering, and labeling distribution, the maximum usable focal density varies among specimens. Adaptive control of the inter-focus lattice spacing is therefore a natural strategy for maintaining reliable computational AO while maximizing acquisition throughput.

Here, we introduce high-throughput aberration-free fluorescence response matrix imaging based on adaptive multifocal illumination. A programmable digital micromirror device (DMD) generates multifocal excitation lattices whose inter-focus spacing is adjusted according to specimen-dependent crosstalk. The acquisition follows an image-scanning geometry, in which the multifocal lattice is scanned across the specimen while fluorescence is recorded with spatially resolved detection. The resulting detector-resolved fluorescence responses are assembled into a two-coordinate fluorescence response matrix. Dual deconvolution then estimates and corrects the excitation and emission OTFs separately. The corrected matrix is combined in the spatial-frequency domain using a SIM-like spectral synthesis rule to reconstruct the aberration-corrected image. We refer to this image-formation step as multifocal virtual SIM (MvSIM) when applied before aberration correction and as adaptive optics-MvSIM (AO-MvSIM) when applied after. Using this adaptive multifocal strategy, we reduced the number of DMD frameshifts and camera frames required for Nyquist acquisition by approximately three orders of magnitude, enabling acquisition of planes with ~$10^6$ illumination positions within seconds to a few minutes. We validated the AO-MvSIM approach by imaging through an approximately 300-μm-thick aberrating layer and applied it to fixed mouse brain and kidney tissue. The accelerated acquisition preserved the AO capability and achieved an estimated lateral resolution of ~100 nm. In biological tissue, the increased acquisition speed enabled high-resolution, depth-resolved volumetric imaging while correcting specimen-dependent aberrations. Together, these results establish adaptive multifocal illumination as a practical route to high-throughput dual deconvolution fluorescence response matrix imaging while preserving its ability to separately correct excitation and emission aberrations.

## Results

### Multifocal acquisition and reconstruction of the fluorescence response matrix

Figure 1 outlines the acquisition and reconstruction sequence. A DMD projects a programmable multifocal lattice onto the specimen, and a scientific complementary metal–oxide–semiconductor (sCMOS) camera records the spatially resolved fluorescence response of every focus (Fig. 1a). Before the full scan, a spacing sweep measures the crosstalk between neighboring responses in the same specimen under the same imaging conditions (Fig. 1b,c). A dense lattice needs fewer frameshifts but produces more crosstalk; a sparse lattice reduces crosstalk at the cost of more frameshifts. The optimal lattice spacing $s_{\mathrm{opt}}$ is the smallest allowed spacing at which the fitted crosstalk is at or below a background-referenced threshold $\tau$ (Methods).

The selected lattice is then translated through one lattice period in one-pixel steps (Fig. 1d). Each illumination position is $\mathbf{r}_{\mathrm{i}} = p\mathbf{q} + \mathbf{s}$, where $p$ is the lattice period in DMD pixels ($p = s_{\mathrm{opt}}/\Delta x$, with $\Delta x$ the scan step at the specimen plane in the same length units as the lattice spacing), $\mathbf{q}$ is the two-dimensional integer index of a focus and $\mathbf{s}$ is the lattice shift shared by all foci in one frameshift. Thus, $\mathbf{r}_{\mathrm{i}}$ is the absolute illumination position on the registered DMD grid, and the physical lattice spacing is $p$ times the projected micromirror pitch. A complete square scan takes $p^2$ frameshifts, each containing an $n \times n$ array of foci, and samples $(np)^2$ illumination positions. Multifocal acquisition therefore reduces the ideal number of frameshifts by a factor of $n^2$ relative to single-focus sampling of the same grid.

For the cell and tissue settings in Table 1, 400–3,025 foci were recorded per frameshift, and about $10^6$ illumination positions were sampled in 12–260 s per plane. These gains are relative to single-focus acquisition of the same detector-resolved fluorescence response matrix.

The frameshift images are registered to the DMD grid, and a local region of detection (ROD) is extracted around each focus. The responses are reindexed by their illumination and detection coordinates to assemble the fluorescence response matrix (Fig. 1e and Supplementary Fig. S1). Each column holds the response for one illumination position $\mathbf{r}_{\mathrm{i}}$, and each row corresponds to one absolute detection position $\mathbf{r}_{\mathrm{d}}$. The element $f(\mathbf{r}_{\mathrm{d}}, \mathbf{r}_{\mathrm{i}})$ thus records the fluorescence detected at $\mathbf{r}_{\mathrm{d}}$ under illumination at $\mathbf{r}_{\mathrm{i}}$, so that each column traces the response across the ROD (Fig. 1e, white curves), and samples with $\mathbf{r}_{\mathrm{d}} = \mathbf{r}_{\mathrm{i}}$ lie on the diagonal. This matrix approximates a single-focus measurement if crosstalk from neighboring foci and truncation by the ROD are small.

Fourier transformation over both coordinate sets gives the spectral fluorescence response matrix. Within each locally shift-invariant patch, dual deconvolution estimates the excitation OTF, the object spectrum and the emission OTF separately (Fig. 1f). After correction, the frequency pairs that satisfy $\mathbf{k} = \mathbf{k}_{\mathrm{d}} + \mathbf{k}_{\mathrm{i}}$ are combined to reconstruct the object. We call spectral synthesis before and after correction MvSIM and AO-MvSIM, respectively. Figure 1g shows a two-color AO-MvSIM image of BPAE cells, with F-actin in green and mitochondria in red.

For comparison, conventional laser-scanning microscopy (LSM) and conventional MSIM images were reconstructed from the same raw data. For conventional LSM, the fluorescence within each ROD was summed without a pinhole and assigned to the corresponding illumination position. For conventional MSIM, each excitation spot was digitally pinholed with a two-dimensional Gaussian mask with a full width at half maximum (FWHM) of 1 Airy unit, contracted by a factor of two about its center and then summed over frameshifts, which is equivalent to pixel reassignment. Conventional LSM and MSIM images were then processed by blind Richardson–Lucy deconvolution, except for the raw frame sums in Fig. 2a,e,i and the conventional MSIM images shown without deconvolution in Fig. 3c and Supplementary Fig. S2a.

**Lattice spacing sets the trade-off between throughput and reconstruction fidelity**

We examined the effect of the lattice spacing in BPAE cells labeled for F-actin and mitochondria. At spacings of 0.9, 1.8 and 2.7 μm, the raw frame sums retain substantial diffuse fluorescence in the actin channel (Fig. 2a,e,i). Conventional MSIM delineates the filament network more clearly (Fig. 2b,f,j), and the mitochondrial images separate neighboring structures better as the spacing increases (Fig. 2c,g,k). The two-color reconstruction at 2.7 μm combines both channels over the same field (Fig. 2m).

Images of neighboring responses and their intensity profiles show progressively deeper valleys as the spacing increases (Fig. 2d,h,l). The measured valley-to-peak crosstalk ratios are about 58%, 8.5% and 1.4% at 0.9, 1.8 and 2.7 μm, respectively (Fig. 2n). At the 90-nm DMD scan step, these spacings correspond to lattice periods of 10, 20 and 30 pixels and require 100, 400 and 900 frameshifts, so lower crosstalk costs more frameshifts. At 2.7 μm, 1,156 foci per frameshift sample 1,040,400 illumination positions in 900 frameshifts, a 1,156-fold reduction relative to single-focus acquisition (Table 1).

For the conventional MSIM mitochondrial images, the Pearson correlation coefficient (PCC) and structural similarity index measure (SSIM) relative to the 2.7-μm dataset increase with spacing (Fig. 2o), from about 0.95 and 0.79 at 0.9 μm to about 0.98 and 0.93 at 1.8 μm. Both metrics equal one at the reference spacing, so this comparison measures agreement with the widest-spacing dataset, not with independent ground truth. The AO-MvSIM reconstructions show the same trend. Correction cannot fully compensate for excessive crosstalk: merged features remain at 0.9 μm, and the 2.7-μm dataset gives the clearest mitochondrial network of the three (Supplementary Fig. S3). We therefore used 2.7 μm for the full acquisition, a conservative choice given that the crosstalk was already below 10% at 1.8 μm.

**Path-resolved aberration correction under multifocal acquisition**

We characterized the lateral response with nominally 100-nm fluorescent beads (Supplementary Fig. S4). Gaussian fits to the ensemble-averaged profiles of ten beads gave apparent FWHMs of 234.0 nm for conventional LSM, 181.0 nm for MvSIM and 144.4 nm for AO-MvSIM, and the mean of the individual-bead widths was 144.5 ± 6.0 nm for AO-MvSIM (mean ± s.d.; Supplementary Table S1). The response narrowed after spectral synthesis and narrowed further after dual deconvolution in the spatial-frequency domain. Subtracting the nominal 100-nm bead diameter in quadrature, as an approximate correction for the bead size, converts the measured 144.4 nm into an effective PSF width of about 100 nm.

We then tested correction through an approximately 300-μm-thick artificial aberrating layer placed above a BPAE-cell specimen and imaged with a 60×, 1.0-NA water-dipping objective (Fig. 3a). The broadened and directionally distorted responses in the raw camera image show the strength of the aberration (Fig. 3b). A lattice spacing of 6.0 μm gave 400 foci per frameshift and 2,500 frameshifts for $10^6$ illumination positions. With a 100-ms exposure, the measured acquisition time was 260 s (Table 1), against an exposure-only lower bound of 100,000 s for single-focus acquisition on the same grid.

The composite overview compares conventional MSIM without deconvolution, AO-MvSIM and MvSIM over the same field (Fig. 3c). In the enlarged region, conventional MSIM after blind Richardson–Lucy deconvolution (10 iterations; Gaussian initial PSF, $\sigma = 2$ pixels) and uncorrected MvSIM leave neighboring mitochondria poorly separated (Fig. 3d,e). AO-MvSIM recovers continuous tubules with dark gaps between them (Fig. 3f), in agreement with the image acquired independently after removal of the aberrating layer (Fig. 3g). Profiles through the marked structures show better-separated peaks after correction (Fig. 3k): AO-MvSIM resolves four distinct peaks along the line, whereas the conventional MSIM and MvSIM profiles fluctuate without clear valleys between them.

The excitation and emission PSFs were obtained separately from the local OTF estimates of dual deconvolution (Fig. 3h,i). Both are elongated and carry a secondary lobe, but the relative lobe strength differs between the two paths (Fig. 3j): along the marked line, both PSFs peak on one side of the illumination position and carry a secondary lobe on the other side. The corresponding local modulation transfer functions (MTFs) and phase transfer functions (PTFs) are shown in Supplementary Fig. S5e–h, and Supplementary Fig. S6 extends the comparison to baseline, mild and stronger aberration.

We quantified the agreement with the layer-removed reference by PCC and SSIM after subpixel registration. Conventional MSIM and uncorrected MvSIM gave PCCs of 0.49 and 0.51 and SSIMs of 0.09 and 0.11; AO-MvSIM raised them to 0.95 and 0.77. Path-resolved correction thus recovers the mitochondrial network under multifocal acquisition.

**Recovery of dendritic spine features in brain tissue**

We next imaged EGFP-labeled neuronal processes in a fixed Thy1-EGFP mouse brain section. Resolving the neck that connects a spine head to its parent dendrite is necessary for assessing spine morphology. The lattice spacing was 1.71 µm at 17 µm depth and 1.90 µm at 22 µm depth (Fig. 4a,c). The response matrices contained 980,100 and 1,000,000 illumination positions and were acquired in 324 frameshifts over 12 s and 400 frameshifts over 15 s, respectively (Table 1), 3,025 and 2,500 times fewer frames than single-focus sampling of the same grids.

At 17 µm depth, conventional LSM shows the dendrite and bright protrusions, but diffuse fluorescence hides several connections between spine heads and the shaft. Conventional MSIM separates some of these features, and AO-MvSIM resolves the narrow connections and suppresses the surrounding haze (Fig. 4a). At profile location L1, the apparent FWHM is 207 nm for conventional LSM and 139 nm for both conventional MSIM and AO-MvSIM. At L2, the widths are 276, 229 and 134 nm, respectively. At L3, AO-MvSIM gives 160 nm, whereas the two comparison methods give no width (n/a) because noise produces multiple peaks across the selected spine neck (Fig. 4b). These apparent widths include the finite size of the labeled structures.

At 22 µm depth, stronger background lowers the signal-to-noise ratio (SNR), and the difference between AO-MvSIM and the comparison methods grows (Fig. 4c). In conventional LSM and MSIM, the protrusions remain embedded in diffuse fluorescence and noise; their profiles contain several peaks of similar height, and a width can be measured only for MSIM at L3 (227 nm; Fig. 4d). AO-MvSIM separates the protrusions and their connections, with apparent widths of 332, 171 and 187 nm at L1–L3. Cyan and magenta arrowheads mark examples of lollipop-type and mushroom-type spines.

Across three fields at 16, 17 and 22 µm depth, a spine-detectability analysis with shared detection parameters found 23, 49, 34 and 62 protrusions in the conventional LSM, conventional MSIM, MvSIM and AO-MvSIM reconstructions, respectively (Supplementary Fig. S7 and Supplementary Table S2). These counts measure relative visibility under a fixed scoring procedure; the dendritic centerlines were traced on the corrected images, and the counts were not validated against independent ground truth.

**Depth-resolved volumetric imaging of kidney tissue**

Finally, we imaged a fixed 16-µm-thick mouse kidney section labeled with Alexa Fluor 488–wheat germ agglutinin (WGA). Imaging through water and a cover glass with a 60×, 1.0-NA objective adds refractive-index transitions to the optical heterogeneity of the tissue (Fig. 5a). With a 6.0-µm lattice spacing and 400 foci per frameshift, $10^6$ illumination positions per plane were sampled in 2,500 frameshifts over 85 s with a 10-ms exposure (Table 1), a 400-fold reduction in frames and an approximately 118-fold reduction in time relative to exposure-only single-focus acquisition.

Volumetric and x–z views show the labeled renal structures across depth (Fig. 5b). Conventional MSIM preserves their overall arrangement, whereas AO-MvSIM gives sharper contours; in the *x–z* views, the WGA-labeled membranes that appear as axially smeared bands in conventional MSIM are confined to narrower axial extents in AO-MvSIM. At the representative depths of 4, 6 and 8 µm, AO-MvSIM sharpens a rounded WGA-positive profile (magenta arrowheads), resolves fine features along luminal boundaries (cyan) and separates the boundaries between adjacent renal structures (yellow; Fig. 5c,e). Conventional MSIM images before and after blind Richardson–Lucy deconvolution are compared with the corresponding AO-MvSIM images in Supplementary Fig. S2.

Maximum-intensity projections (MIPs) of the conventional MSIM and AO-MvSIM depth series are shown in Fig. 5d and f. The series comprised ten planes at 1-µm axial intervals, the deepest at 10 µm, and each MIP keeps the maximum intensity along the depth axis at each lateral position. With a 10-ms exposure per frameshift, diffuse background in the conventional MSIM projection obscures boundaries and fine features; the AO-MvSIM projection from the same raw data suppresses this haze and delineates WGA-positive contours and submicrometer structures. The individual planes and the projections together show depth-resolved acquisition and correction of response matrices in kidney tissue.

## Discussion

### Practical acceleration of dual deconvolution

Multifocal acquisition reduces the number of frameshifts needed for dual deconvolution by measuring many illumination positions at once. In the cell and tissue datasets, 400–3,025 foci were recorded per frameshift, and about $10^6$ illumination positions were sampled in 12–260 s per plane. These gains are relative to single-focus acquisition of the same response matrix; they are not an equivalent speed advantage over conventional multifocal methods that reconstruct an image directly. Because the illumination and detection coordinates are retained, the excitation and emission OTFs can still be estimated separately: parallel acquisition cuts the number of frameshifts without discarding the measurements needed for path-resolved correction.

### Relationship to MISM and MvSIM image formation

The multifocal geometry builds on MSIM and MISM[15,16,17], in which scattering and crosstalk limit the usable lattice density[18,19]. For dual deconvolution, crosstalk additionally breaks the association between illumination and detection coordinates. The spacing sweep measures how crosstalk falls with spacing in each specimen, but the threshold $\tau$ alone does not fix the best spacing: in the BPAE dataset, the more conservative 2.7-μm lattice improved reconstruction fidelity even though the measured crosstalk was already below 10% at 1.8 μm.

The reconstruction keeps the sampled fluorescence response matrix so that the excitation and emission paths are corrected separately before spectral synthesis. This differs from forming a pixel-reassigned image and then deconvolving it. The frequency-sum relation itself is shared with the structured-illumination description of ISM[23,24]; what is different here is the use of the two-coordinate matrix for path-resolved correction. Other developments in ISM include rapid nonlinear imaging[25] and super-resolution fluorescence lifetime imaging[26]. Focus-ISM[27] and structured detection[28] improve optical sectioning by suppressing out-of-focus fluorescence.

### Aberration correction under accelerated multifocal acquisition

The aberrating-layer experiment tests whether parallel acquisition retains enough information for computational aberration correction. AO-MvSIM recovered mitochondrial structures that matched the independently acquired layer-removed reference, and the estimated excitation and emission PSFs showed distinct distortions (Fig. 3). The bead measurements gave an apparent FWHM of 144.4 nm after correction (Supplementary Fig. S4); because the beads have a finite size, this width characterizes the measured response rather than the PSF alone.

The correction also held at low SNR. At 22 μm depth in brain tissue, AO-MvSIM resolved spine protrusions at all three profile locations, whereas conventional LSM gave no measurable width and conventional MSIM gave one only at L3 (Fig. 4c,d). In the kidney series, the corrected projection showed sharper contours than the conventional MSIM projection from the same raw data (Fig. 5d,f). Within each local patch, dual deconvolution estimates the excitation and emission responses jointly from many detector-resolved measurements, so the correction draws on spatial information shared across those measurements.

Unlike hardware AO with deformable mirrors or spatial light modulators[7,8], this approach estimates the correction computationally from the fluorescence measurements themselves. Other matrix-based and computational methods also recover aberrations from measured images[11,12,29,30]. Hardware and computational correction can be combined: hardware AO can remove large aberrations before acquisition, and response matrix processing can then estimate residual or field-dependent aberrations. Computational correction cannot, however, restore spatial frequencies that have fallen below the noise floor of the measurement.

### Volumetric biological imaging enabled by accelerated acquisition

The kidney depth series shows why acquisition time matters for response matrix imaging. Matrices were collected at ten planes and corrected with local excitation and emission estimates at each depth. MSIM and related methods already support volumetric imaging[15,18,19]; the present depth series additionally retains the measurements needed for path-resolved correction. The volumetric views, individual planes and projections in Fig. 5 show improved structural visibility but do not provide an independent measurement of axial resolution.

### Current limitations in acquisition and reconstruction

Elapsed acquisition time is limited by pattern transfer and camera readout. In the present implementation, binary patterns were streamed from the host computer as 8-bit image buffers, and the frameshift period was 34–39 ms for the datasets acquired with 10-ms exposures. Preloading the pattern sequence and faster hardware playback would reduce this overhead, within the limits set by camera readout and exposure.

Matrix reconstruction also requires dense illumination sampling and a detection region large enough to retain the aberrated response. CPU-based correction took about 60 s per $25 \times 25$ μm² patch for five iterations. GPU-based structured illumination microscopy (SIM) reconstruction[31] and accelerated iterative reconstruction[32] offer routes to reducing this cost. Noise control and regularization also matter at low signal levels[33,34,35,36], although here they must preserve the separate excitation and emission estimates. The frameshift comparison also assumes matched per-focus power and exposure; the gain attainable at fixed total illumination power depends on the photon budget.

### Toward local and real-time adaptive aberration-free imaging

The present implementation selects one lattice spacing before each scan. Measurements at several positions or depths could instead set local spacings, and adaptive scanning and event-driven acquisition in other microscopy platforms[37,38] suggest feedback strategies. Such control would avoid unnecessary sampling in sparse regions while keeping adequate separation where labeling, scattering or aberrations are stronger.

A DMD and a camera added to an epifluorescence microscope therefore provide a route to path-resolved computational aberration correction. Faster pattern delivery, camera readout and reconstruction would further shorten the time per plane and make volumetric imaging more practical.

## Methods

### Experimental setup

The DMD-based multifocal fluorescence response matrix imaging system (Supplementary Fig. S8) used two excitation lasers: a 488-nm Cobolt 06-MLD for EGFP, Alexa Fluor 488-labeled specimens and yellow-green fluorescent beads, and a 561-nm OBIS LS 561-150 for MitoTracker Red.

A digital micromirror device (DMD; DLI4120VIS-951080P, Digital Light Innovations, with a Texas Instruments DMD) generated the multifocal lattice and scanned it as a two-dimensional raster scanner in place of galvanometric mirrors. The lattice patterns were streamed from the host computer as 8-bit image buffers.

Excitation and emission were separated by a multiband dichroic beamsplitter. Fluorescence was collected through a 60× water-dipping objective (NA 1.0; N60X-NIR, Nikon) or a 100× silicone oil-immersion objective (NA 1.35; CFI SR

HP Plan Apochromat Lambda S 100XC Sil, Nikon), filtered by a multiband emission filter (ZET488/561m-TRF, Chroma) and detected by an sCMOS camera (pco.edge 4.2, PCO AG).

**Crosstalk measurement and lattice-spacing selection**

We quantified the crosstalk between neighboring focal responses with the crosstalk ratio (CR). For each lattice spacing $s$, CR was calculated from the valley and the higher peak of an ensemble-averaged, dark-corrected line profile across two neighboring responses:

$$\mathrm{CR}(s) = \frac{\bar{I}_{\mathrm{valley}}(s)}{\bar{I}_{\mathrm{peak}}(s)} \times 100\%. \tag{1}$$

In Eq. (1), the numerator is the minimum between the two peaks, the denominator is the higher of the two peaks and overbars denote ensemble averaging, so CR expresses the valley intensity as a percentage of the peak intensity.

The crosstalk threshold is set relative to the residual background floor $C$, which contains the detector dark signal and diffuse fluorescence: detector noise dominates when the fluorescence background is negligible, whereas tissue background can dominate at depth. The threshold is expressed in the same percentage units as CR:

$$\tau = C + k\sigma_{\mathrm{bg}}. \tag{2}$$

Here $C$ is the residual offset of the fitted crosstalk curve in Eq. (5), $\sigma_{\mathrm{bg}}$ is the standard deviation of the background floor in CR percentage points, and $k$ sets how far above the floor the fitted crosstalk may rise; $k$ was chosen according to the background fluctuations of each acquisition condition. The criterion therefore requires the fitted excess of CR above the floor to be no larger than $k\sigma_{\mathrm{bg}}$.

The optimal lattice spacing is the smallest allowed spacing at or above the fitted threshold crossing:

$$s_{\mathrm{opt}} = \min\{s \in S_{\mathrm{DMD}} : s \geq s_\tau\}, \text{ where } \widehat{\mathrm{CR}}(s_\tau) = \tau. \tag{3}$$

In Eq. (3), $S_{\mathrm{DMD}}$ is the set of allowed lattice spacings, $\tau$ is the threshold of Eq. (2) and $s_\tau$ is the spacing at which the fitted curve crosses $\tau$. Because the fitted curve decreases with spacing, rounding up to the next allowed spacing keeps the predicted CR at or below $\tau$. If no allowed spacing satisfies the condition, the criterion cannot be met within the available range.

For the detector offset, dark frames were acquired without excitation at the same camera settings and exposure time as the fluorescence measurements. We averaged these frames, took the minimum of the averaged dark image as a single offset and subtracted it from every pixel of each fluorescence image:

$$I_{\mathrm{corr}}(\mathbf{r}, s) = I_{\mathrm{raw}}(\mathbf{r}, s) - D_{\mathrm{off}},$$
$$D_{\mathrm{off}} = \min_{\mathbf{r}'} \langle I_{\mathrm{dark}}(\mathbf{r}') \rangle, \quad \langle I_{\mathrm{dark}}(\mathbf{r}) \rangle = \frac{1}{N_{\mathrm{d}}} \sum_{k=1}^{N_{\mathrm{d}}} I_{\mathrm{dark},k}(\mathbf{r}). \tag{4}$$

Here $I_{\mathrm{raw}}(\mathbf{r}, s)$ is the raw fluorescence image at lattice spacing $s$, $I_{\mathrm{dark},k}(\mathbf{r})$ is the $k$th of $N_{\mathrm{d}}$ dark frames and $\langle I_{\mathrm{dark}}(\mathbf{r}) \rangle$ is their average. Subtracting the minimum $D_{\mathrm{off}}$ removes a uniform offset but keeps the spatial variation of the dark image, so the residual dark contribution to $I_{\mathrm{corr}}(\mathbf{r}, s)$ is non-negative at every pixel. We did not subtract the maximum dark level $D_{\max}$ or the local profile minimum; the residual dark signal and the diffuse fluorescence therefore remain in the valley and enter the criterion through $C$ and $\sigma_{\mathrm{bg}}$ in Eq. (2). Distinguishing detector offsets from pixel-dependent read noise is relevant to sCMOS imaging[39]; noise accumulation in frame-summed measurements is discussed in the Supplementary Information.

Before each full acquisition, the multifocal lattice was expanded along ±*x* and ±*y* in the most densely labeled region of the specimen. At each spacing, profiles along the lines joining neighboring peaks were extracted from the dark-corrected image, aligned and averaged, and CR was calculated with Eq. (1). This measurement captures nearest-neighbor crosstalk and does not separately quantify the contributions of more distant foci.

The BPAE spacing sweep illustrates the measured trend (Fig. 2n and Supplementary Fig. S3d): CR was about 58%, 8.5% and 1.4% at lattice spacings of 0.9, 1.8 and 2.7 μm, corresponding to DMD periods of 10, 20 and 30 micromirrors. For this comparison we used a 10% reference level for $\tau$; the ratio at 1.8 μm is below it, and 2.7 μm gives the lowest measured CR. The 10% level is specific to the BPAE comparison and is not a universal threshold for tissue with a different background. The profiles in Fig. 2d,h,l and Supplementary Fig. S3d show the corresponding decrease of the normalized valley intensity.

We describe the spacing dependence $\widehat{\mathrm{CR}}(s)$ by an empirical single-exponential decay with a constant offset, fitted over the spacing range measured for each specimen (0.9–2.7 μm for the BPAE sweep):

$$\widehat{\mathrm{CR}}(s) = A \exp\left(-\frac{s}{s_\mathrm{c}}\right) + C, \tag{5}$$

where $s$ is the lattice spacing, $A$ is the amplitude (%), $s_\mathrm{c}$ is the characteristic decay length in the same units as $s$ and $C$ is the residual offset (%). The BPAE ratios fall to about 0.15 and 0.16 of the preceding value over successive 0.9-μm increments, consistent with an approximately exponential decrease over this range. The offset summarizes the residual floor over the fitted range; any spacing-dependent background remains part of the measured ratio. The model is empirical and does not identify the physical origin of the decay. The 100% marker at zero spacing in Fig. 2n is excluded from the fit.

The fitted curve defines the crossing $s_\tau$, which Eq. (3) maps to an allowed spacing in $S_\mathrm{DMD}$. Only even lattice periods were used: the ROD is cropped symmetrically around the 2 × 2 illumination spot with a margin $\delta_\rho = (\rho - 2)/2$, so its width $\rho$ must be even, and we set $\rho = p$ so that the RODs tile each lattice cell. Although the DMD can generate odd periods, this convention restricts $S_\mathrm{DMD}$ to spacings two micromirrors apart. Rounding up from $s_\tau$ therefore selects the densest allowed lattice whose fitted CR does not exceed $\tau$. The threshold and the crosstalk curve depend on the specimen and the imaging condition, and once $s_\mathrm{opt}$ is selected the lattice spacing stays fixed for the full scan.

For the BPAE comparison, the spacing sweep used the MitoTracker Red channel with 561-nm excitation. We used the 2.7-μm lattice for the full acquisition because it gave the lowest measured CR and the clearest reconstruction of the tested spacings; it corresponds to a 30-micromirror period and 900 frameshifts and serves as the reference in Fig. 2o. This is a conservative choice: the threshold rule alone does not determine the spacing that maximizes reconstruction quality.

**Fluorescence response matrix acquisition by multifocal illumination**

For detector positions within the ROD assigned to each illumination position, the fluorescence response is reconstructed from the frameshift images as

$$f^{(\rho)}(\mathbf{r}_\mathrm{d},\mathbf{r}_\mathrm{i}) = \left[W\{g^\mathrm{cam}_{k(\mathbf{r}_\mathrm{i})}\}(\mathbf{r}_\mathrm{d}) - b\right]_+, \tag{6}$$

where $g^\mathrm{cam}_k$ is the frameshift image acquired at lattice shift $k$, with the frame index determined by the shift at which $\mathbf{r}_\mathrm{i}$ is illuminated. The operator $W$ registers camera images to the DMD grid with subpixel accuracy, and $\mathbf{r}_\mathrm{i}$ and $\mathbf{r}_\mathrm{d}$ are expressed in pixels on this common grid. The scalar $b$ is the estimated background, $[x]_+ = \max(x, 0)$ denotes non-negative clipping and the superscript $\rho$ denotes a $\rho \times \rho$ ROD. Illumination positions follow $\mathbf{r}_\mathrm{i} = p\mathbf{q} + \mathbf{s}$, where $p$ is the lattice period in pixels, $\mathbf{q} \in \mathbb{Z}^2$ indexes a focus within the lattice and $\mathbf{s} \in \{0,1,\dots,p-1\}^2$ is the lattice shift shared by all foci in one frameshift. In a *y*-fast raster scan with one-pixel steps, the $p^2$ frameshifts step through $\mathbf{s}$ starting from $\mathbf{s} = \mathbf{0}$, so that each focus sweeps the $p \times p$ cell anchored at $p\mathbf{q}$ (Supplementary Fig. S1). The lattice period, number of foci, ROD width and scan-grid dimensions of the datasets in Figs. 2, 3 and 5 are listed in Supplementary Table S3.

Reindexing the responses by illumination and detection coordinates assembles the fluorescence response matrix of Fig. 1e. Each illumination position is sampled once, and the ROD recorded there fills the corresponding column. The ROD pixels lie at absolute detection coordinates $\mathbf{r}_\mathrm{d} = \mathbf{r}_\mathrm{i} + \Delta\mathbf{r}$, where $\Delta\mathbf{r}$ spans the offsets within the $\rho \times \rho$ ROD, and each pixel maps to the row indexed by $\mathbf{r}_\mathrm{d}$. The element in row $\mathbf{r}_\mathrm{d}$ and column $\mathbf{r}_\mathrm{i}$ thus records the fluorescence detected at $\mathbf{r}_\mathrm{d}$ under illumination at $\mathbf{r}_\mathrm{i}$, and it is populated only where $\mathbf{r}_\mathrm{d} - \mathbf{r}_\mathrm{i}$ falls within the ROD window. Because both coordinates share the same grid, the zero-offset samples $(\Delta\mathbf{r} = \mathbf{0}, \text{i.e.}, \mathbf{r}_\mathrm{d} = \mathbf{r}_\mathrm{i})$ lie on the diagonal.

The spatial-domain fluorescence response matrix is $\mathbf{F}^{(\mathrm{r})}$, with elements

$$[\mathbf{F}^{(\mathrm{r})}]_{mn} = f^{(\rho)}(\mathbf{r}_{\mathrm{d},m}, \mathbf{r}_{\mathrm{i},n}), \tag{7}$$

where $m$ and $n$ enumerate the vectorized detection and illumination coordinates. Fourier transformation over both two-dimensional coordinates yields the spectral fluorescence response matrix $\mathbf{F}^{(\mathrm{k})}$.

**Reference-based image quality assessment**

**Pearson correlation coefficient (PCC):**

$$\mathrm{PCC} = \frac{\sum_{\mathbf{r}} [g(\mathbf{r}) - \overline{g}][u(\mathbf{r}) - \overline{u}]}{\sqrt{\sum_{\mathbf{r}} [g(\mathbf{r}) - \overline{g}]^2 \ \sum_{\mathbf{r}} [u(\mathbf{r}) - \overline{u}]^2}}. \tag{8}$$

For the spacing comparison in Fig. 2o, the reference $g$ is the conventional MSIM mitochondrial reconstruction from the 2.7-μm dataset and $u$ is the reconstruction at the compared spacing; for the aberrating-layer comparison in Fig. 3, the reference is the independently acquired layer-removed image. Images were registered by maximizing the PCC over scale and subpixel translation and intensity-matched by least squares over the common overlapping region. Overbars denote means over that region, sums run over its pixels, and PCC and SSIM were both evaluated over it.

**Structural similarity index measure (SSIM)[40]:**

$$\mathrm{SSIM} = \frac{1}{N_\mathrm{w}} \sum_{\mathbf{r} \in \Omega_\mathrm{w}} \frac{(2\mu_g\mu_u + C_1)(2\sigma_{gu} + C_2)}{({\mu_g}^2 + {\mu_u}^2 + C_1)(\sigma_g^2 + \sigma_u^2 + C_2)}. \tag{9}$$

SSIM was calculated from the local means $\mu_g$ and $\mu_u$, variances $\sigma_g^2$ and $\sigma_u^2$, and covariance $\sigma_{gu}$ at each window center, using a unit-sum 11 × 11 Gaussian window (standard deviation, 1.5 pixels), $C_1 = (0.01L)^2$ and $C_2 = (0.03L)^2$, where $L$ is the intensity dynamic range ($L = 1$ for images normalized to [0, 1]). $\Omega_\mathrm{w}$ is the set of evaluated window centers $\mathbf{r}$, and $N_\mathrm{w}$ is their number.

**Acquisition parameters and timing**

**Table 1 | Acquisition parameters and comparison with single-focus acquisition**

| Fig. | Specimen and condition | Objective | Lattice spacing (μm) | DMD scan step (nm) | Frameshifts | Exposure (ms) | Acquisition (s) |
|---|---|---|---|---|---|---|---|
| S4 | Fluorescent beads | 100×, 1.35 NA (Sil) | 0.54 | 90 | 36 | 5 | 10 |
| 2 | BPAE cells, two-color | 100×, 1.35 NA (Sil) | 2.7 | 90 | 900 | 10 | 35 |
| 3 | BPAE cells, aberrating layer | 60×, 1.0 NA (W) | 6.0 | 120 | 2,500 | 100 | 260 |
| 4 | Mouse brain tissue (17 μm depth) | 60×, 1.0 NA (W) | 1.71 | 95 | 324 | 10 | 12 |
| 4 | Mouse brain tissue (22 μm depth) | 60×, 1.0 NA (W) | 1.90 | 95 | 400 | 10 | 15 |
| 5 | Mouse kidney tissue | 60×, 1.0 NA (W) | 6.0 | 120 | 2,500 | 10 | 85 |

| Dataset | Foci per frameshift | Frameshifts | Single-focus frames | Frame gain | Time gain |
|---|---|---|---|---|---|
| Beads (Supplementary Fig. S4) | 27,556 | 36 | 992,016 | 27,556× | 496× |
| BPAE (Fig. 2) | 1,156 | 900 | 1,040,400 | 1,156× | 297× |
| Aberrating layer (Fig. 3) | 400 | 2,500 | 1,000,000 | 400× | 385× |
| Brain, 17 μm (Fig. 4) | 3,025 | 324 | 980,100 | 3,025× | 817× |
| Brain, 22 μm (Fig. 4) | 2,500 | 400 | 1,000,000 | 2,500× | 667× |
| Kidney (Fig. 5) | 400 | 2,500 | 1,000,000 | 400× | 118× |

Sil, silicone oil; W, water. Acquisition time is the measured elapsed time; values for Fig. 2 are per channel and values for Fig. 5 are per plane. Total exposure times were 0.18 s for the beads, 9 s per channel for Fig. 2, 250 s for Fig. 3, 3.24 and 4 s for the 17- and 22-μm brain datasets, and 25 s for Fig. 5; the remaining time was spent on host-to-DMD transfer of the 8-bit pattern buffers and camera readout. A different relay magnification gave the 95-nm scan step in Fig. 4. Single-focus frame counts use the same illumination grid with one frame per position. Frame gain is the single-focus frame count divided by the number of frameshifts. Counts assume complete square lattices and omit edge masking and calibration frames. Time gain is the single-focus exposure-only time divided by the measured multifocal acquisition time, assuming matched per-focus power and exposure.

**Computational performance of matrix reconstruction**

On an Intel Core i9-11900K workstation with 128 GB of RAM, CPU-only matrix reconstruction took 14, 43 and 95 s for the Fig. 2 datasets acquired at lattice spacings of 0.9, 1.8 and 2.7 μm, respectively.

**Computational performance of dual deconvolution correction**

CPU-only dual deconvolution of the nine patches of the mouse kidney dataset (Alexa Fluor 488 channel) in Fig. 5 took 550 s in total for a 75 × 75 μm² region divided into 25 × 25 μm² patches, that is, about 60 s per patch for five iterations (about 12 s per iteration).

**Preparation of fluorescent beads**

A 10% poly-L-lysine solution diluted with distilled water (1 mL) was dispensed onto a glass slide (P000BMBS, Marienfeld Superior) and covered with a cover glass (P000BMCU, Marienfeld Superior) for 30 s to spread it evenly over the surface. After drying for 30 min, the coated slide was rinsed with distilled water. A bead suspension was prepared by mixing 100 μL of yellow-green fluorescent carboxylate-modified polystyrene microspheres with a nominal diameter of 100 nm (FluoSpheres Carboxylate-Modified Microspheres, 0.1 μm, excitation/emission maxima 505/515 nm; F8803, Invitrogen, Thermo Fisher Scientific) with 5–10 mL of distilled water, and a volume chosen to give sparsely distributed beads in the imaging region was applied to the coated slide. After the beads had adsorbed to the poly-L-lysine layer for 10 min, the slide was rinsed with distilled water and air-dried.

**Preparation of the artificial aberrating layer sample**

A fully cured polydimethylsiloxane (PDMS) layer about 300 μm thick was placed on a high-precision cover glass (0107052, Paul Marienfeld). A cleaning polymer solution (First Contact, FCDFR) was applied over the PDMS layer and, before it had dried completely, its surface was roughened with sandpaper; the scratching was varied to produce surface structures ranging from slowly varying, grating-like patterns to rapidly varying random patterns. As the polymer dried at room temperature, irregular surface curvatures developed and produced spatially varying wavefront errors and scattering.

Immersion oil (N6468400, Olympus; $n = 1.518$) was placed between the commercial BPAE slide and the aberrating layer for refractive-index matching, and the layer was placed over the specimen to add aberrations during imaging.

**Preparation of the BPAE cells**

A commercially prepared FluoCells Prepared Slide #1 (F36924, Invitrogen, Thermo Fisher Scientific) with fixed, multiply labeled bovine pulmonary artery endothelial cells was used as supplied, without additional fixation or staining. Mitochondria had been labeled in living cells with MitoTracker Red CMXRos before fixation; after fixation and permeabilization, F-actin had been stained with Alexa Fluor 488 phalloidin and nuclei with DAPI. For two-color imaging, Alexa Fluor 488 phalloidin and MitoTracker Red CMXRos were excited at 488 and 561 nm, respectively.

**Preparation of the mouse brain tissue**

Twelve-week-old adult Thy1-EGFP line M mice (stock #007788, The Jackson Laboratory) were anesthetized by intraperitoneal injection of ketamine (100 mg/kg) and xylazine (10 mg/kg) and euthanized by decapitation. The brains were removed and fixed in 4% paraformaldehyde (PFA) at 4 °C overnight, washed three times in phosphate-buffered saline (PBS), cryoprotected in 30% (w/v) sucrose in PBS at 4 °C until equilibrated, embedded in optimal cutting temperature (OCT) compound, frozen and sectioned at a thickness of 30–50 μm using a cryostat at about −20 °C. Sections were collected on positively charged glass slides and stored at −80 °C. Before imaging, they were washed three times in PBS, secured in a plastic imaging dish and immersed in PBS.

**Preparation of the mouse kidney tissue**

A commercially prepared FluoCells Prepared Slide #3 (F24630, Invitrogen, Thermo Fisher Scientific) containing a 16-μm-thick cryostat section of mouse kidney was imaged as supplied. Elements of the glomeruli and convoluted tubules were labeled with Alexa Fluor 488-conjugated wheat germ agglutinin, filamentous actin in the glomeruli and brush border with Alexa Fluor 568 phalloidin and nuclei with DAPI, and the section was mounted in ProLong antifade medium.

All experimental procedures involving animals were approved by the Korea University Institutional Animal Care and Use Committee (approval number KUIACUC-2025-0002).

**Acknowledgments**

This work was supported by the National Research Foundation of Korea (NRF) grants [No. RS-2023-00213310 (S.Y.), No. RS-2024-00442818 (S.L., T.L. and W.C.), No. RS-2025-24132968 (J.H.H.)] and the Institute of Information & Communications Technology Planning & Evaluation (IITP) grant (No. RS-2025-25464788, Development of High-Sensitivity, High-Resolution Quantum MRI Technology Based on Diamond NV Centers, 20% (S.L., T.L. and W.C.)) funded by the Korean government (MSIT).

**Author contributions**

S.L., S.Y. and W.C. conceived the project. S.L., S.P. and T.L. constructed the experimental setup. S.L. and S.Y. developed computational algorithms. S.L. performed the experiments and acquired the datasets. With guidance from W.C., S.L. analyzed the data. S.L. and J.H.H. prepared and provided biological specimens. S.L., S.Y. and W.C. prepared the manuscript. W.C. supervised the project.

**Code availability**

The custom acquisition and analysis code is available from the corresponding authors upon reasonable request.

## Data availability

The supporting datasets are available from the corresponding authors upon reasonable request.

## Competing interests

The authors declare no competing interests.

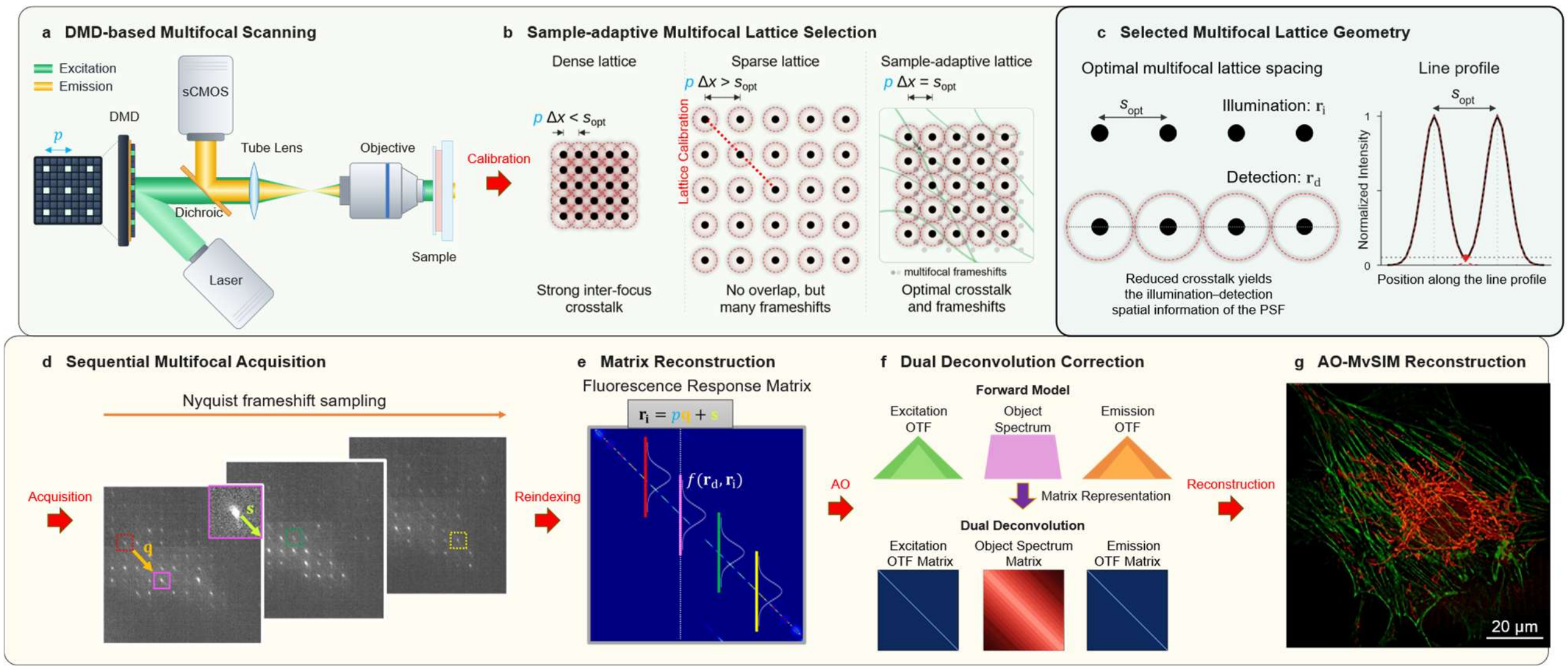

**Fig. 1 | Sample-adaptive multifocal acquisition and path-resolved reconstruction. a**, A DMD projects a multifocal lattice onto the specimen, and an sCMOS camera records the fluorescence responses. The lattice period $p$ is measured in DMD pixels. **b**, A preliminary spacing sweep balances crosstalk against the number of frameshifts required for a full scan; the projected micromirror pitch converts the DMD lattice period into the lattice spacing at the specimen plane. **c**, The optimal lattice spacing $s_{\mathrm{opt}}$ separates neighboring responses while retaining their illumination and detection coordinates; the valley-to-peak ratio of the line profile quantifies the crosstalk. **d**, Sequential lattice shifts sample the illumination grid. The vector $\mathbf{q}$ indexes a focus within the lattice, and $\mathbf{s}$ is the lattice shift shared by all foci in one frameshift. Colored boxes mark four regions of detection (RODs) whose responses appear as the correspondingly colored columns in **e**. **e**, Reindexing assigns each response to the illumination coordinate $\mathbf{r}_{\mathrm{i}} = p\mathbf{q} + \mathbf{s}$ and the absolute detection coordinate $\mathbf{r}_{\mathrm{d}}$. Columns and rows of the fluorescence response matrix index $\mathbf{r}_{\mathrm{i}}$ and $\mathbf{r}_{\mathrm{d}}$, respectively; each column holds the response $f(\mathbf{r}_{\mathrm{d}}, \mathbf{r}_{\mathrm{i}})$ detected across the ROD for one illumination position (white curves), and zero-offset samples lie on the diagonal. Coordinates are expressed on the registered DMD grid, and the physical lattice spacing is $p$ times the projected micromirror pitch. **f**, Fourier transformation and dual deconvolution separate the excitation OTF, object spectrum and emission OTF within each local patch, and the corrected frequency pairs are combined by spectral synthesis. **g**, Two-color AO-MvSIM reconstruction of BPAE cells, with F-actin in green and mitochondria in red. Scale bar, 20 µm.

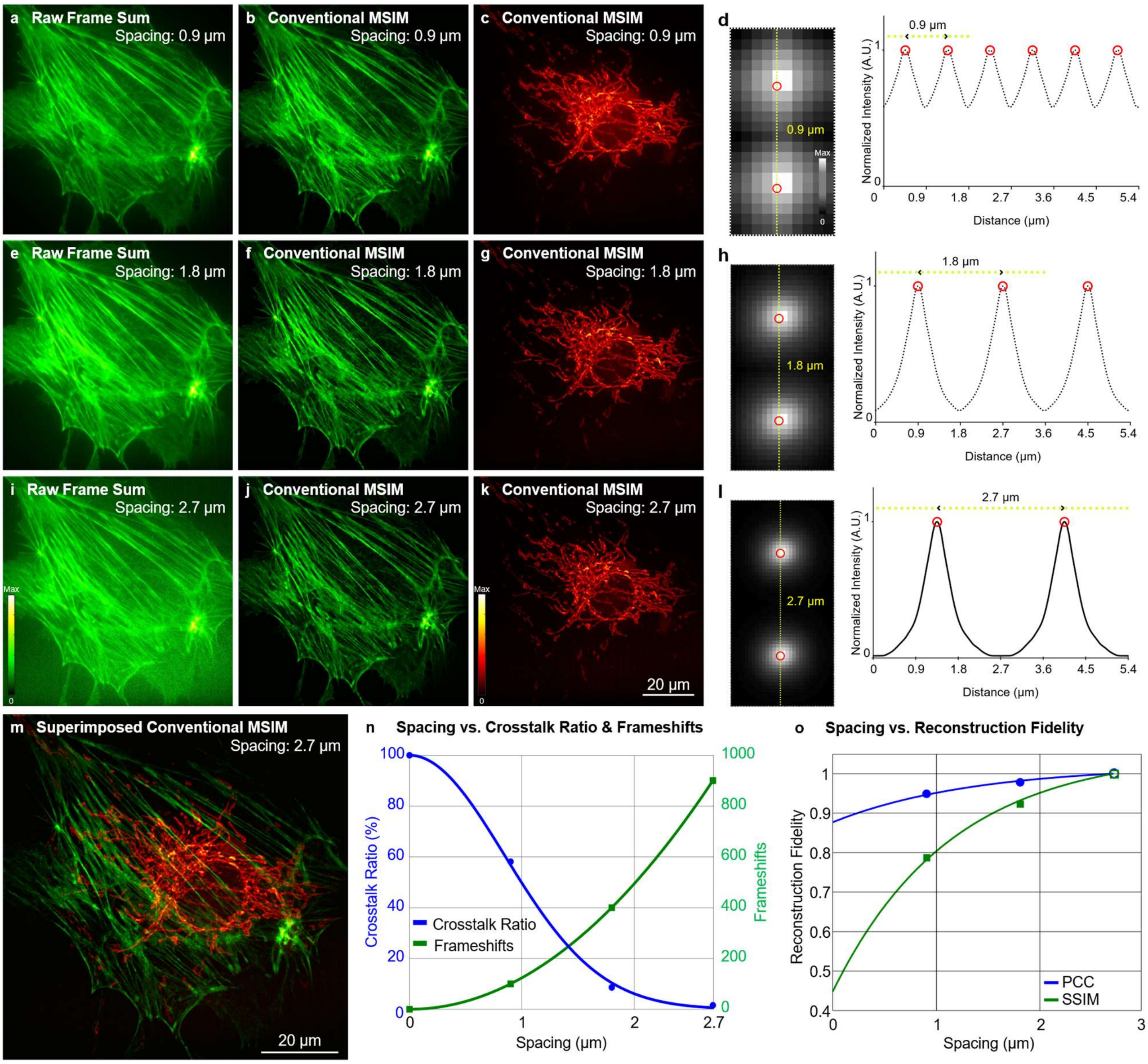

**Fig. 2 | Lattice spacing controls crosstalk and reconstruction fidelity.** BPAE cells labeled with Alexa Fluor 488 phalloidin (F-actin) and MitoTracker Red (mitochondria), imaged at lattice spacings of 0.9, 1.8 and 2.7 µm (top to bottom). **a**,**e**,**i**, Raw frame sums in the actin channel (sum over all frameshifts). **b**,**f**,**j**, Conventional MSIM reconstructions of F-actin. **c**,**g**,**k**, Conventional MSIM reconstructions of mitochondria. **d**,**h**,**l**, Neighboring fluorescence responses and normalized intensity profiles at the three spacings; yellow lines mark the profile positions and red circles the response peaks. **m**, Superimposed conventional MSIM reconstruction at 2.7 µm, with F-actin in green and mitochondria in red. **n**, Crosstalk ratio (blue, left axis) and required number of frameshifts (green, right axis) versus lattice spacing. The measured ratios are about 58%, 8.5% and 1.4%, corresponding to 100, 400 and 900 frameshifts; markers at nonzero spacings denote the measured conditions. The blue curve is a visual guide, not the fitted single-exponential model of Eq. (5), and the 100% point at zero spacing is a reference outside the measured range. **o**, PCC (blue) and SSIM (green) of the conventional MSIM mitochondrial images in **c**,**g**,**k**, using **k** (2.7 µm) as the reference; both metrics equal one at the reference spacing by definition. Scale bars, 20 µm in **k** (for **a**–**c**, **e**–**g** and **i**–**k**) and **m**.

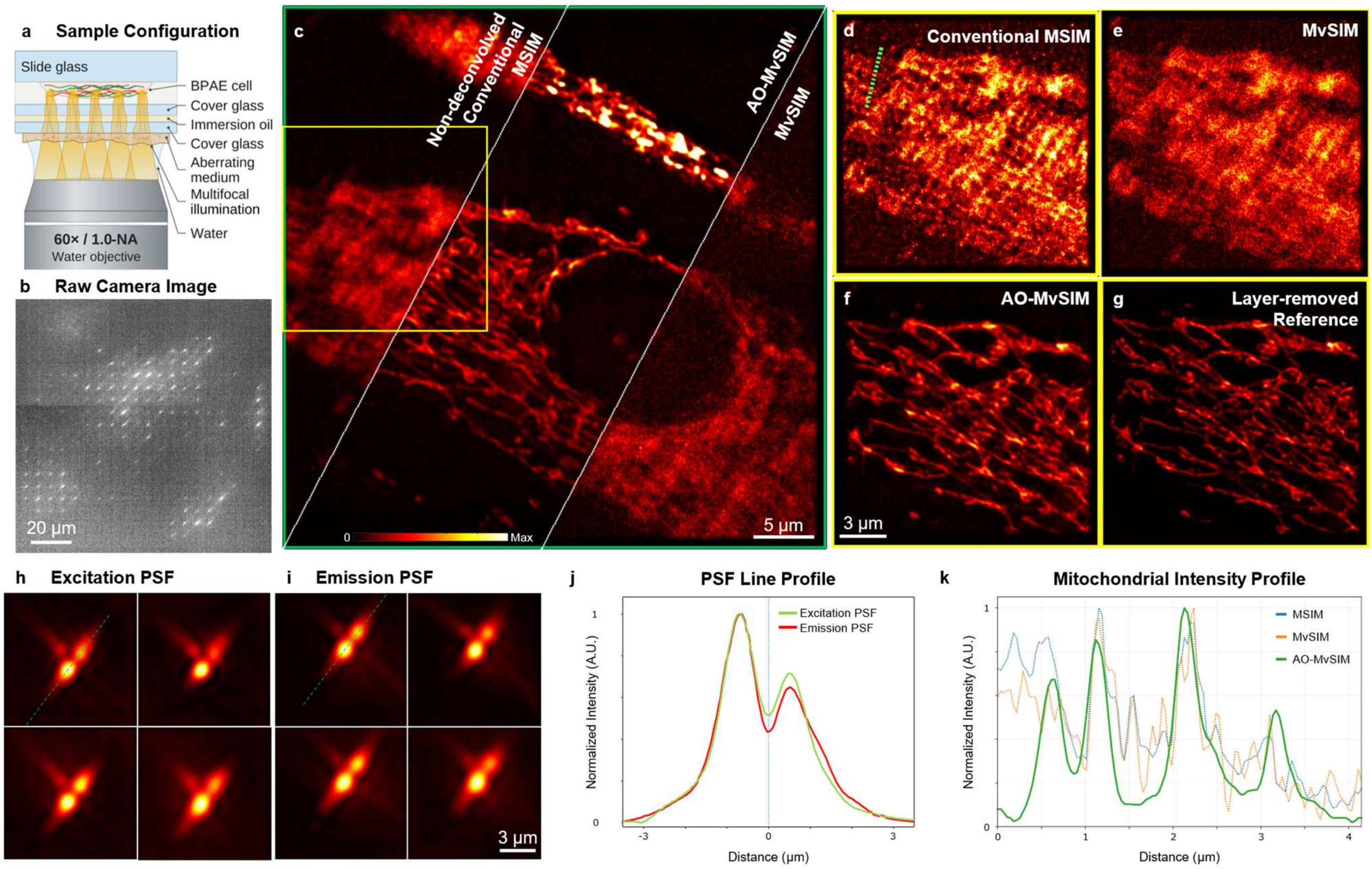

**Fig. 3 | Path-resolved correction through an artificial aberrating layer.** **a**, BPAE-cell imaging through an approximately 300-µm-thick artificial aberrating layer with a 60×, 1.0-NA water-dipping objective. **b**, Raw camera image of the multifocal fluorescence responses. **c**, Composite overview of conventional MSIM without deconvolution (left), AO-MvSIM (center) and MvSIM (right), separated by white diagonal lines; the yellow box marks the region enlarged in **d**–**g**. **d**–**g**, Conventional MSIM after blind Richardson–Lucy deconvolution (**d**; 10 iterations, Gaussian initial PSF with $\sigma$ = 2 pixels), MvSIM (**e**), AO-MvSIM (**f**) and the independently acquired layer-removed reference (**g**). **h**,**i**, Excitation (**h**) and emission (**i**) PSFs recovered for four local regions, arranged in corresponding 2 × 2 arrays. **j**, Normalized profiles along the green dashed lines in the upper-left PSFs; green and red denote the excitation and emission paths, and the blue vertical dotted line marks the illumination position. **k**, Normalized intensity profiles through corresponding mitochondrial structures along the line marked in **d**, comparing conventional MSIM (blue), MvSIM (orange) and AO-MvSIM (green). Fluorescence images show the MitoTracker Red channel. Scale bars, 20 µm (**b**), 5 µm (**c**) and 3 µm (**d**–**i**).

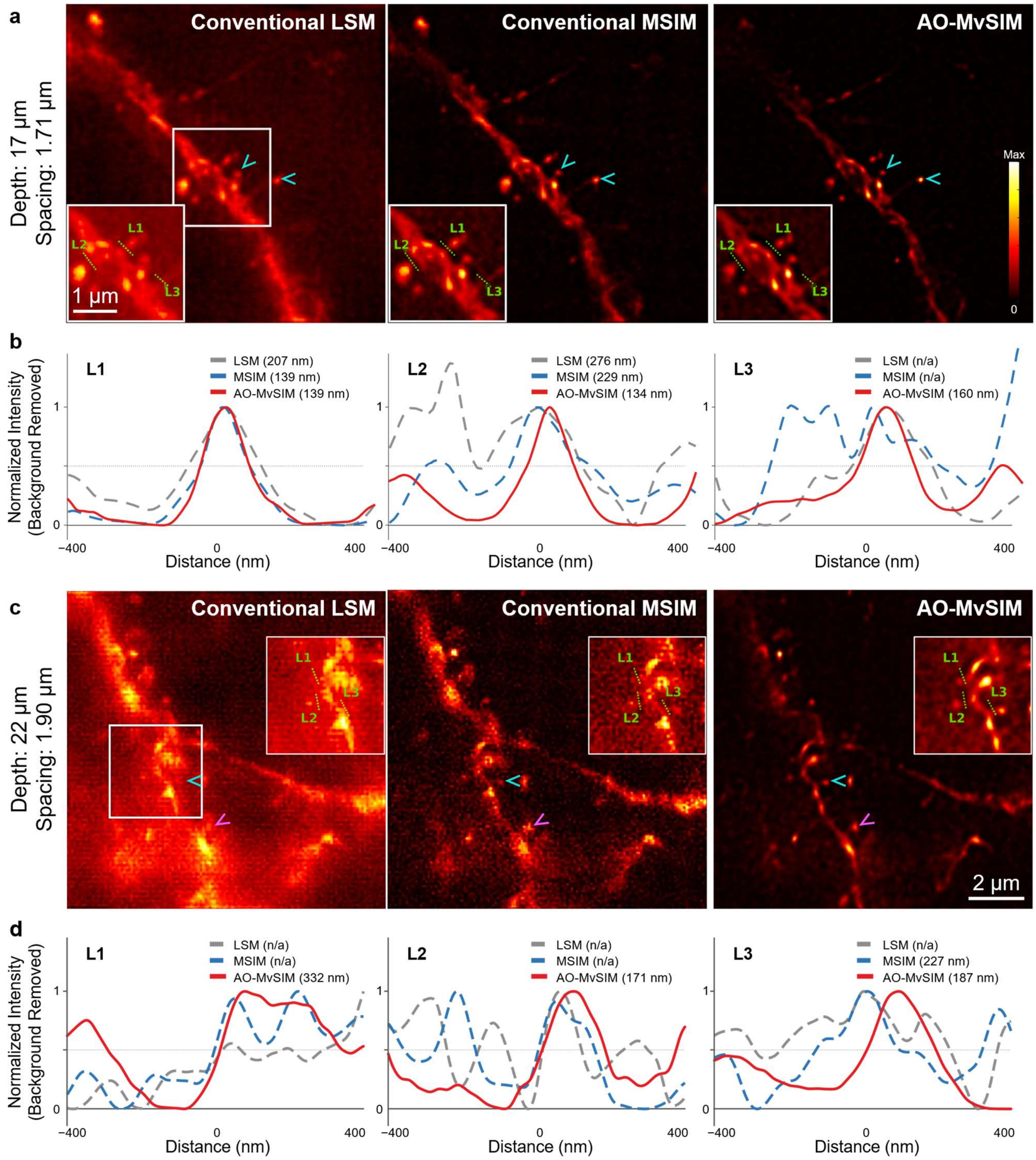

**Fig. 4 | Recovery of dendritic spine features in mouse brain tissue.** EGFP-labeled neuronal processes in a fixed Thy1-EGFP mouse brain section. **a**, Images at 17 µm depth with a lattice spacing of 1.71 µm, reconstructed by conventional LSM (left), conventional MSIM (center) and AO-MvSIM (right). Insets enlarge the white-boxed region and mark profile locations L1–L3 across selected spine necks. **b**, Corresponding background-subtracted profiles, each normalized to the peak of the selected feature. Gray dashed, blue dashed and solid red curves denote conventional LSM, conventional MSIM and AO-MvSIM, respectively. Apparent FWHMs are 207/139/139 nm at L1, 276/229/134 nm at L2 and n/a/n/a/160 nm at L3, in that method order. **c**, Corresponding images at 22 µm depth with a lattice spacing of 1.90 µm. **d**, Profiles at L1–L3 in **c**, with apparent FWHMs of n/a/n/a/332 nm, n/a/n/a/171 nm and n/a/227/187 nm, respectively. A width is reported as not available (n/a) when multiple noise-related peaks prevent a reliable measurement of the selected spine neck. Widths include the finite size of the fluorescent structures. Horizontal dotted lines mark half maximum. Cyan and magenta open arrowheads mark examples of lollipop-type and mushroom-type spines, respectively. Scale bars, 1 µm in the leftmost inset of **a** and 2 µm in the rightmost panel of **c**.

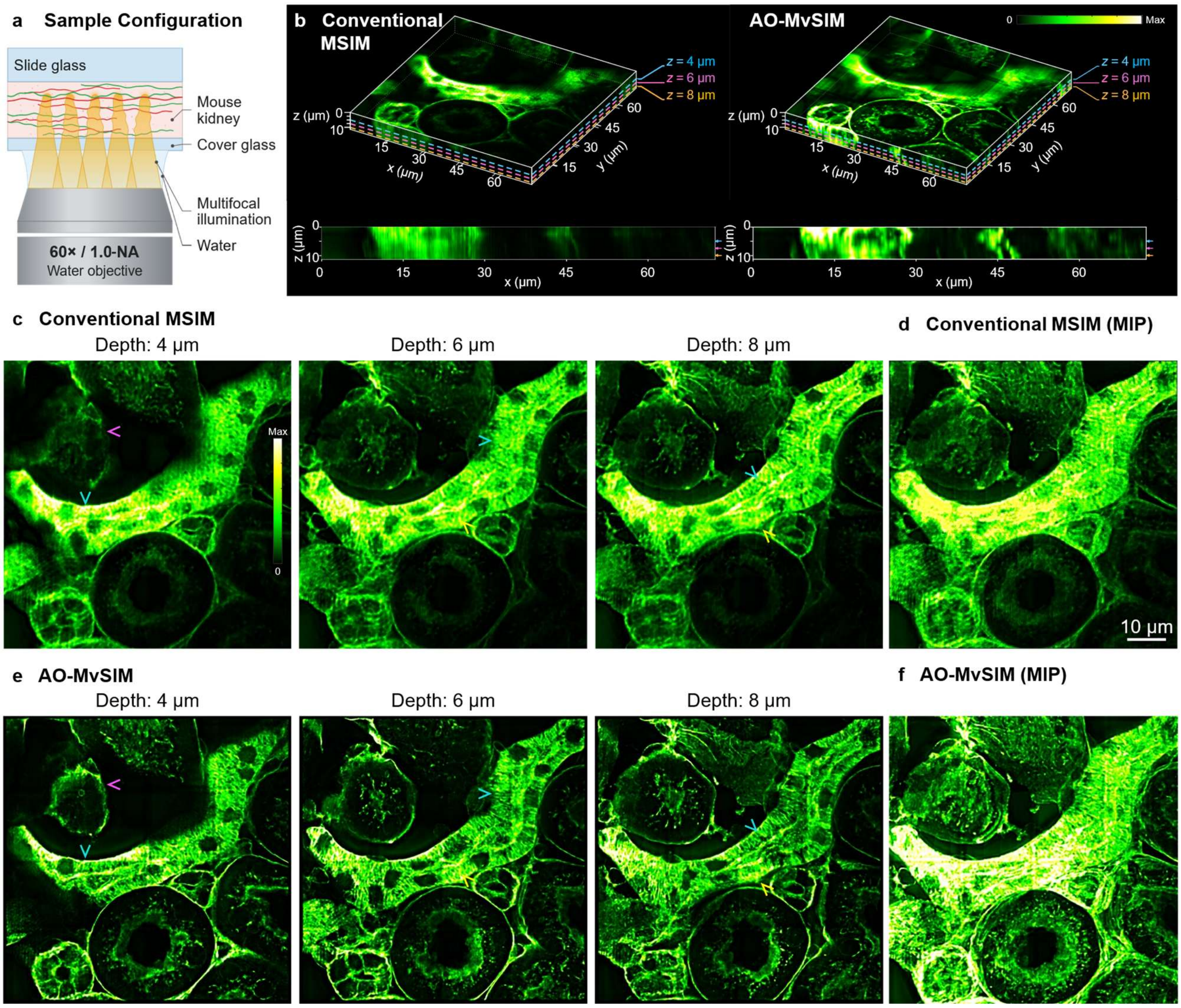

**Fig. 5 | Depth-resolved imaging of mouse kidney tissue.** A fixed 16-µm-thick mouse kidney section labeled with Alexa Fluor 488–wheat germ agglutinin (WGA), imaged with a 6.0-µm lattice spacing. **a**, Imaging geometry with a 60×, 1.0-NA water-dipping objective. **b**, Volumetric reconstructions by conventional MSIM (left) and AO-MvSIM (right), with the corresponding *x*–*z* views below. Colored guides mark depths of 4 µm (cyan), 6 µm (magenta) and 8 µm (orange). **c**, Conventional MSIM images at these depths, from left to right. **d**, Maximum-intensity projection (MIP) of the depth series reconstructed by conventional MSIM; the maximum intensity along the depth axis is kept at each lateral position. **e**, AO-MvSIM images of the planes shown in **c**. Matching open arrowheads mark a rounded WGA-positive profile (magenta), fine features along luminal boundaries (cyan) and boundaries between adjacent renal structures (yellow). **f**, MIP of the AO-MvSIM depth series from the same planes as **d**. The individual planes show depth-specific structure, whereas the MIPs summarize fluorescence across the stack. Scale bar, 10 µm in **d** for the planar images.

# Dual deconvolution multifocal fluorescence response matrix microscopy for aberration-free high-throughput super-resolution imaging
# -Supplementary Information-

Sumin Lim[1*], Sanghyeon Park[1], Tran Thinh Le[1], Jin Hee Hong[1], Seokchan Yoon[2*], Wonshik Choi[1*]

[1]*Department of Physics, Korea University, Seoul 02855, Korea*

[2]*School of Biomedical Convergence Engineering, Pusan National University, Yangsan 50612, Korea*

*Correspondence: lsm9333@naver.com, sc.yoon@pusan.ac.kr, wonshik@korea.ac.kr

**Reconstruction of conventional LSM and MSIM images and fluorescence response matrices**

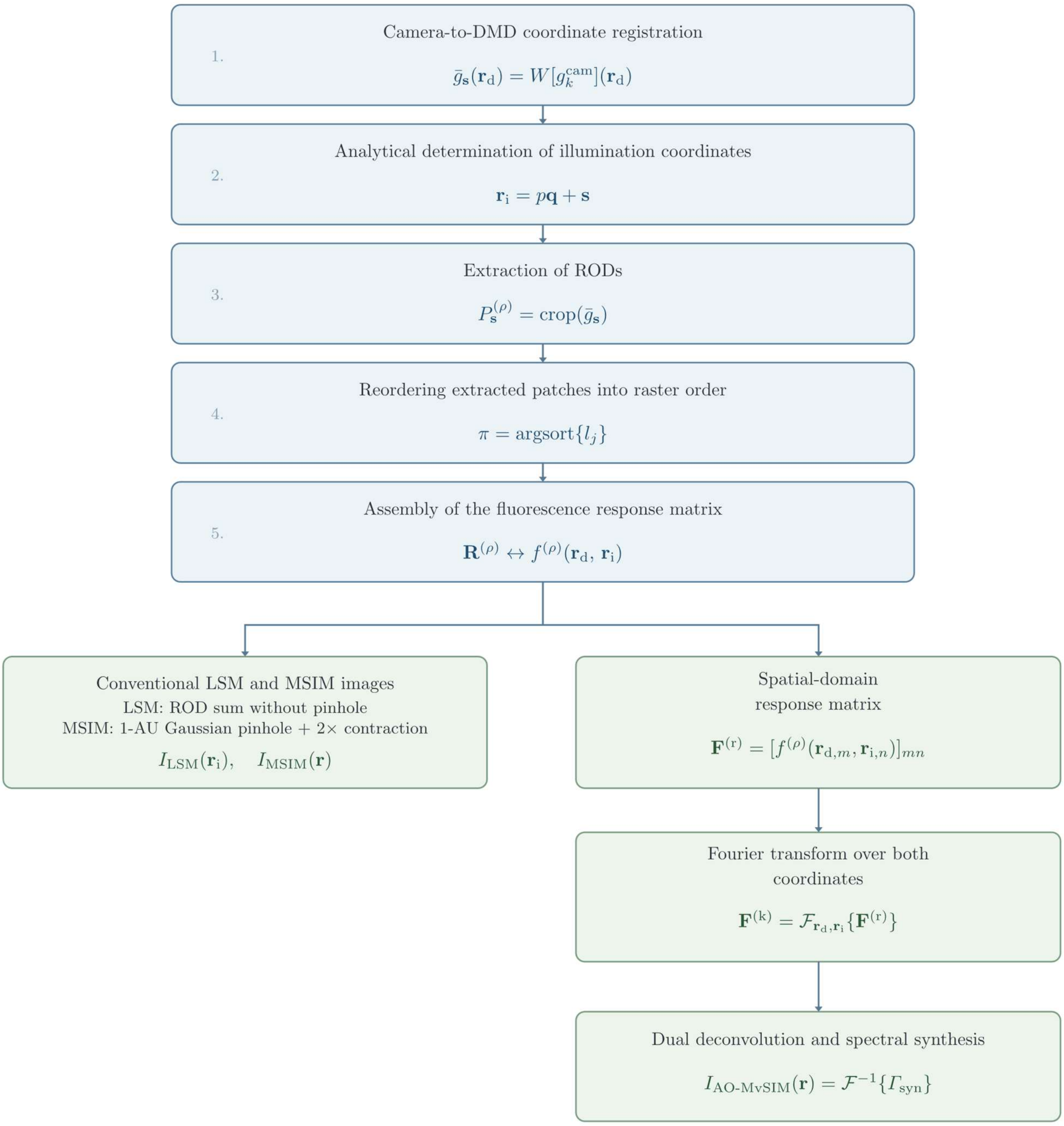

**Fig. S1 | Reconstruction of conventional LSM and MSIM images and fluorescence response matrices.** Steps 1–5 show camera-to-DMD registration, determination of the illumination coordinates, extraction of the RODs, sorting into raster order and assembly of the detector-resolved responses. The left branch forms the conventional LSM and conventional MSIM images indexed by illumination position: conventional LSM sums each ROD without a pinhole, whereas conventional MSIM applies a Gaussian digital pinhole (FWHM, 1 Airy unit) and contracts each response by a factor of two about its center before summation. The right branch retains the absolute illumination and detection coordinates; Fourier transformation over both coordinate sets gives the spectral response matrix, and dual deconvolution and spectral synthesis yield the AO-MvSIM image. The protocol below defines the coordinates and indexing.

**Step 1 | Camera-to-DMD coordinate registration:**

$$\tilde{\mathbf{x}}_\mathrm{D} = \mathbf{T}_\mathrm{aff}\tilde{\mathbf{x}}_\mathrm{C}, \quad \tilde{\mathbf{x}}_\mathrm{D}{}' = \mathbf{T}_\mathrm{fine}\tilde{\mathbf{x}}_\mathrm{D}, \quad \mathbf{T} = \mathbf{T}_\mathrm{fine}\mathbf{T}_\mathrm{aff}. \tag{S1}$$

$\tilde{\mathbf{x}}_\mathrm{C}$ and $\tilde{\mathbf{x}}_\mathrm{D}$ are homogeneous pixel coordinates in the camera and coarsely registered DMD frames, and $\tilde{\mathbf{x}}_\mathrm{D}{}'$ is the refined DMD coordinate. The matrices $\mathbf{T}_\mathrm{aff}$ and $\mathbf{T}_\mathrm{fine}$ are the coarse affine and fine registration transforms; only their composition $\mathbf{T}$ is used to resample the image. The warp operator $W$ includes this single-pass resampling and the origin correction of the local region of interest (ROI).

$$\bar{g}_{\mathbf{s}}(\mathbf{r}_{\mathrm{d}}) = W[g_k{}^{\mathrm{cam}}](\mathbf{r}_{\mathrm{d}}),\ k = s_{\mathrm{y}} + p s_{\mathrm{x}}. \tag{S2}$$

The detector coordinate $\mathbf{r}_{\mathrm{d}}$ is expressed in pixels on the registered DMD grid, and $k$ is the zero-based index of the frameshift.

**Step 2 | Analytical determination of illumination coordinates:**

$$p = m + 2,\ N = np,\ N_{\mathrm{frame}} = p^2. \tag{S3}$$

The 2 × 2 illumination patch and the $m$ intervening dark pixels define the lattice period $p$. Here $n$ is the number of foci per axis and $N$ is the width of the scan grid in DMD pixels; $N$ is unrelated to the acquisition counts in the comparison below.

$$\mathbf{r}_{\mathrm{i}} = p\mathbf{q} + \mathbf{s},\ \mathbf{q} = \left\lfloor \frac{\mathbf{r}_{\mathrm{i}}}{p} \right\rfloor,\ \mathbf{s} = \mathbf{r}_{\mathrm{i}} \bmod p. \tag{S4}$$

Floor and modulo are applied component-wise: $\mathbf{q}$ ranges from 0 to $n - 1$ and $\mathbf{s}$ from 0 to $p - 1$ in each direction, which gives a unique illumination coordinate $\mathbf{r}_{\mathrm{i}}$.

**Step 3 | Extraction of RODs:**

$$P_{\mathbf{s}}{}^{(\rho)}(\mathbf{q}, \boldsymbol{\mu}) = \bar{g}_{\mathbf{s}}(\mathbf{r}_{\mathrm{i}} - \delta_\rho \mathbf{1} + \boldsymbol{\mu}),\ \delta_\rho = \frac{\rho - 2}{2}. \tag{S5}$$

Here, $\boldsymbol{\mu} \in \{0, \dots, \rho - 1\}^2$ is the local ROD coordinate and $\mathbf{1} = (1,1)$. The absolute detection coordinate is $\mathbf{r}_{\mathrm{d}} = \mathbf{r}_{\mathrm{i}} - \delta_\rho \mathbf{1} + \boldsymbol{\mu}$, and the ROD width is $\rho$.

**Step 4 | Reordering extracted patches into raster order:**

$$j = (q_{\mathrm{y}} + n q_{\mathrm{x}}) + n^2(s_{\mathrm{y}} + p s_{\mathrm{x}}),\ l_j = r_{\mathrm{i,y}} + N r_{\mathrm{i,x}}. \tag{S6}$$

$j$ is the multifocal acquisition order and $l_j$ is the column-major raster index of the patch stored at acquisition position $j$; both range from 0 to $N^2 - 1$.

$$\pi = \operatorname{argsort} \{l_j\}_{j=0}^{N^2-1}. \tag{S7}$$

$\pi_l$ is the acquisition index of the patch at raster position $l$. In practice we sorted by the linear index of the padded image, which gives the same order as $l_j$.

**Step 5 | Fluorescence response assembly:** A scalar background $b$ was estimated from the dimmest 50% of the peripheral intensities of the ensemble-averaged $\rho \times \rho$ patches. Subtraction followed by non-negative clipping gave $f^{(\rho)}(\mathbf{r}_{\mathrm{d}}, \mathbf{r}_{\mathrm{i}})$. Summing each ROD without a pinhole gave the conventional LSM image, and summing after Gaussian digital pinholing and twofold contraction gave the conventional MSIM image. Reindexing the retained ROD samples by their absolute detection and illumination coordinates gave $\mathbf{F}^{(\mathrm{r})}$ for dual deconvolution.

**Blind Richardson–Lucy deconvolution of conventional MSIM compared with AO-MvSIM in mouse kidney tissue**

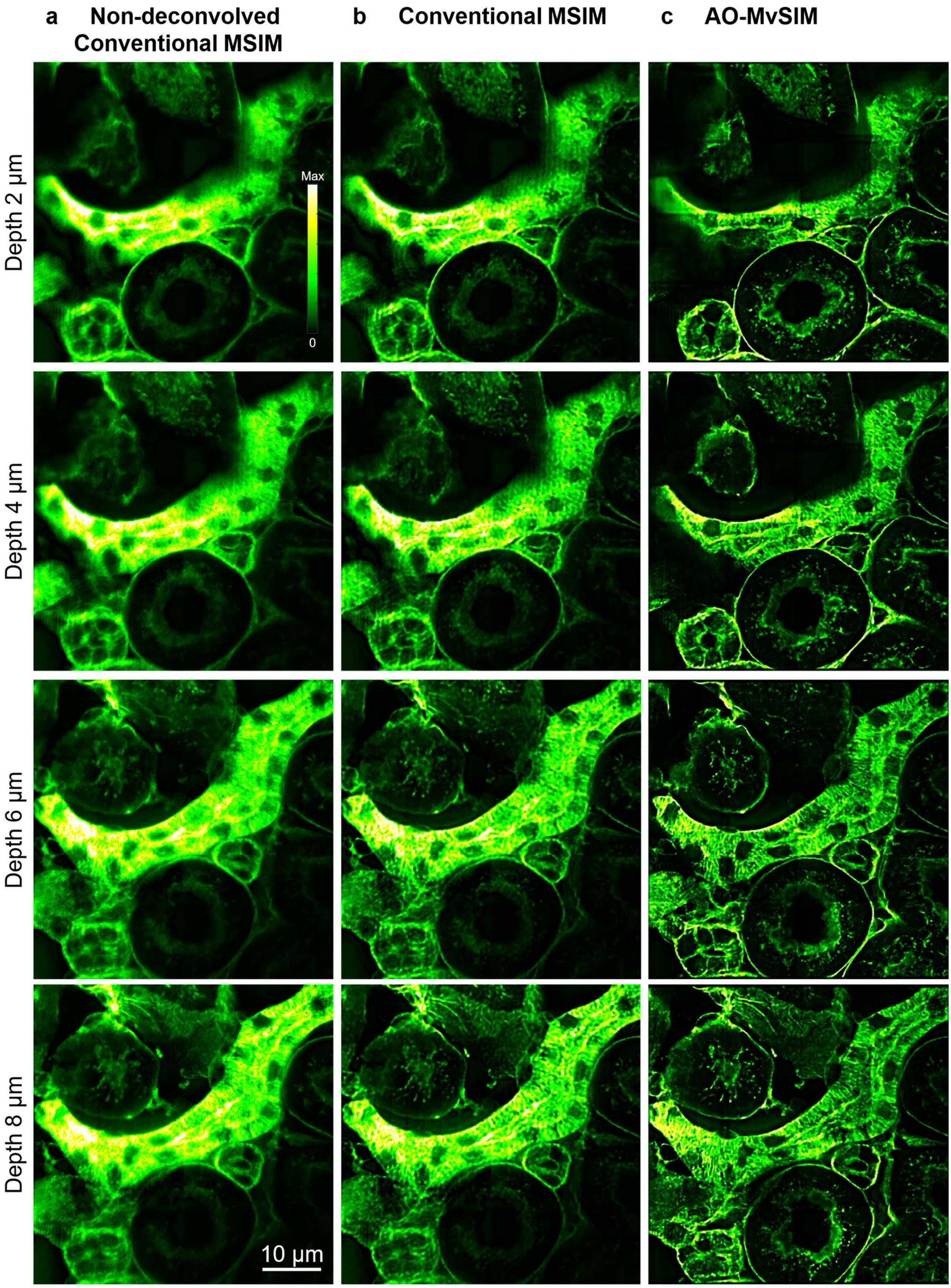

**Fig. S2 | Blind Richardson–Lucy deconvolution of conventional MSIM compared with AO-MvSIM in mouse kidney tissue.** Alexa Fluor 488–wheat germ agglutinin images of the fixed 16-μm-thick mouse kidney section of Fig. 5, acquired with a 6.0-μm lattice spacing. Rows correspond to imaging depths of 2, 4, 6 and 8 μm, from top to bottom. Columns show conventional MSIM without deconvolution (**a**, left), the same images after blind Richardson–Lucy deconvolution (**b**; Gaussian initial PSF with $\sigma = 2$ pixels, 10 iterations) and AO-MvSIM (**c**, right), all reconstructed from the same depth-specific multifocal datasets. Under these settings, blind deconvolution sharpens some boundaries but leaves diffuse background and adds grainy texture, whereas AO-MvSIM resolves continuous WGA-positive contours and fine luminal features. In conventional MSIM, pixel reassignment combines the responses before single-image deconvolution, and aberrations reduce the high-frequency content retained in the combined image; dual deconvolution corrects the excitation and emission paths before spectral synthesis. Each image is normalized to its maximum. Scale bar, 10 μm in the bottom-left panel.

**Dependence of reconstruction quality on lattice spacing**

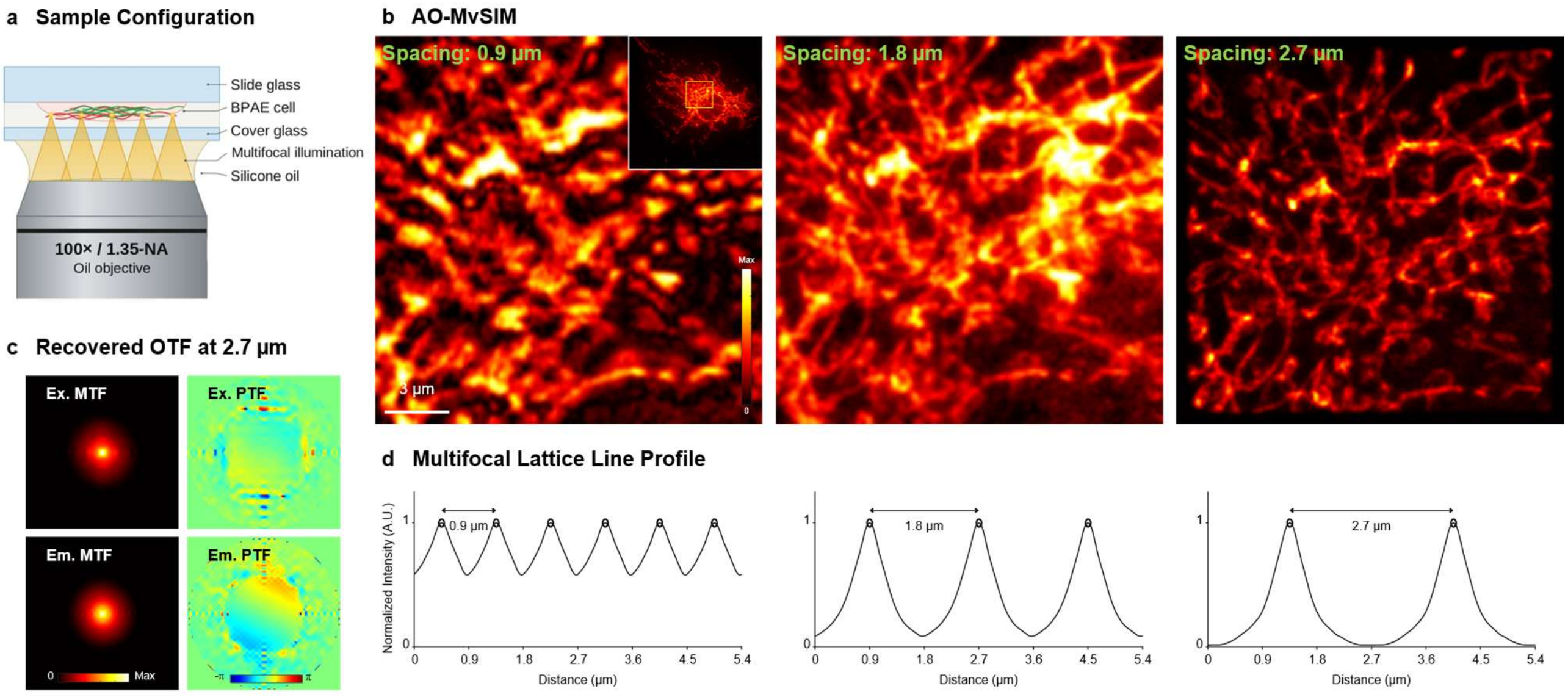

**Fig. S3 | Effect of lattice spacing on reconstruction and crosstalk.** **a**, BPAE-cell imaging configuration with a 100×, 1.35-NA silicone oil-immersion objective. **b**, AO-MvSIM images of the same mitochondrial region at lattice spacings of 0.9, 1.8 and 2.7 µm, from left to right; the inset locates the enlarged region. **c**, Excitation (upper row) and emission (lower row) MTFs (left) and PTFs (right) recovered from the 2.7-µm dataset. The modulation transfer function (MTF) and phase transfer function (PTF) are the OTF magnitude and phase, respectively; phase is displayed from −π to π. **d**, Normalized multifocal intensity profiles at the three spacings; deeper valleys indicate less crosstalk. Scale bar, 3 µm in **b**.

Supplementary Fig. S3 complements the spacing comparison of Fig. 2 with AO-MvSIM reconstructions, recovered OTFs and profiles of neighboring responses. Fluorescence entering the ROD from adjacent foci can degrade the reconstruction even when the central response remains recognizable. The 2.7-µm dataset is the widest-spacing reference in this comparison; larger spacings were not evaluated.

**Lateral resolution characterization using fluorescent beads**

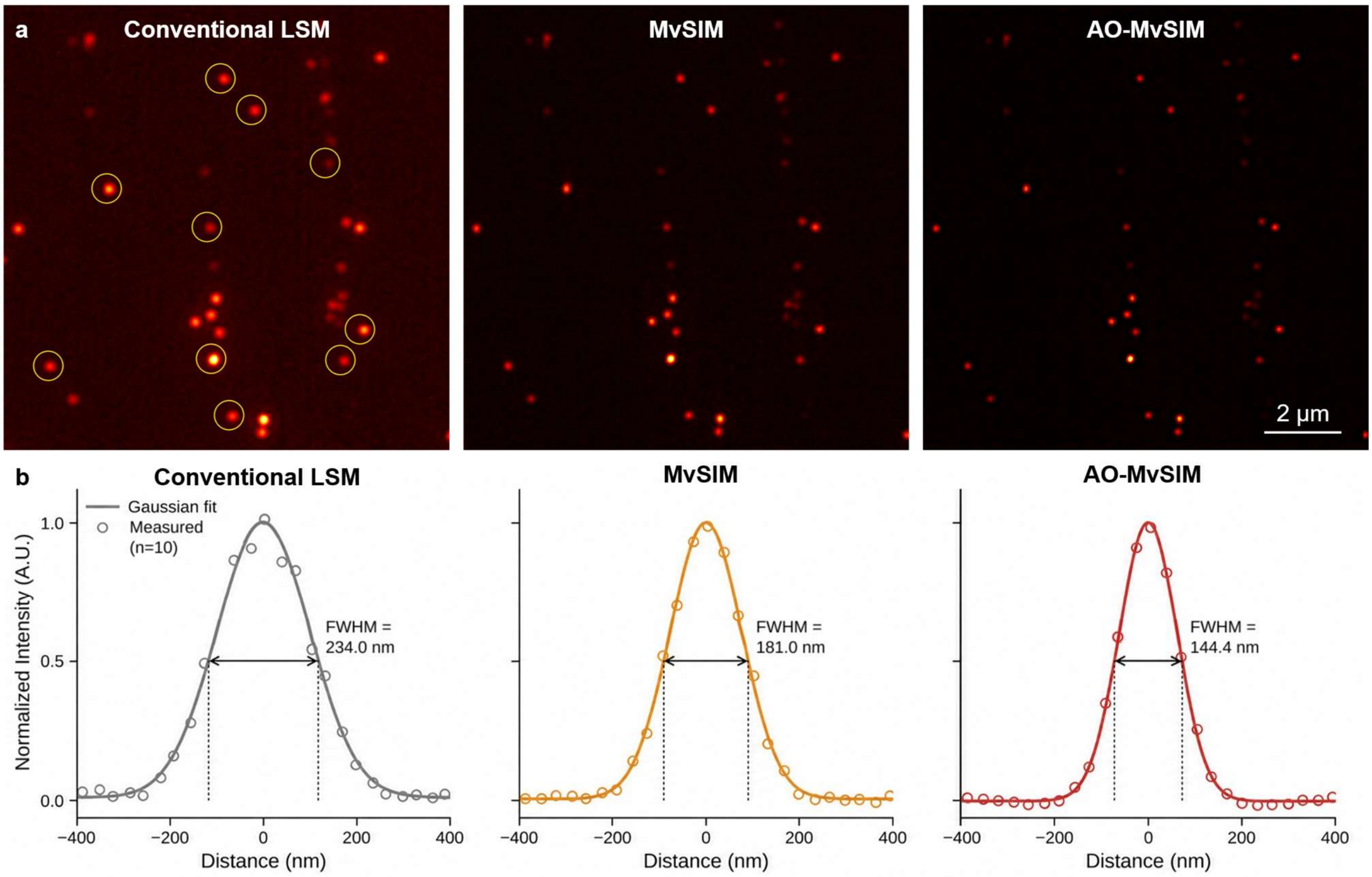

**Fig. S4 | Lateral fluorescence-response widths measured with nominally 100-nm beads. a**, Conventional LSM (left), MvSIM (center) and AO-MvSIM (right) images of yellow-green fluorescent beads excited at 488 nm and imaged with the 1.35-NA objective. Yellow circles in the conventional LSM image mark the ten beads used for analysis. **b**, Subpixel-aligned, normalized lateral profiles averaged over the ten beads (open circles) and Gaussian fits (solid curves). The fitted ensemble-profile FWHMs are 234.0 nm for conventional LSM, 181.0 nm for MvSIM and 144.4 nm for AO-MvSIM; the corresponding means of the individual-bead FWHMs are 237.7 ± 4.5, 182.2 ± 3.6 and 144.5 ± 6.0 nm (mean ± s.d.; $n$ = 10 beads; Table S1). Individual-bead widths are the geometric means of the fitted widths along the two principal axes. All values include the finite bead size. Scale bar, 2 μm in **a**.

**Table S1 | Lateral FWHM of individual fluorescent beads measured with conventional LSM, uncorrected MvSIM and AO-MvSIM**

| Bead | Conventional LSM (nm) | MvSIM, uncorrected (nm) | AO-MvSIM (nm) |
|---|---|---|---|
| 1 | 237.22 | 176.37 | 142.05 |
| 2 | 237.46 | 179.85 | 140.48 |
| 3 | 237.58 | 177.03 | 144.79 |
| 4 | 236.08 | 183.13 | 141.44 |
| 5 | 238.04 | 186.66 | 138.59 |
| 6 | 231.48 | 185.03 | 148.98 |
| 7 | 238.32 | 183.86 | 144.38 |
| 8 | 243.41 | 183.26 | 143.52 |
| 9 | 231.20 | 180.62 | 159.59 |
| 10 | 245.99 | 185.70 | 140.94 |
| **Mean ± s.d.** | **237.7 ± 4.5** | **182.2 ± 3.6** | **144.5 ± 6.0** |

The same ten beads were measured in all three images (FWHM as defined in Fig. S4). The mean FWHM decreases by a factor of 1.30 from conventional LSM to uncorrected MvSIM and by 1.26 from uncorrected MvSIM to AO-MvSIM, corresponding to an overall reduction by a factor of 1.65.

**Acquisition-count comparison at matched scan step**

The throughput reference in the main text is single-focus acquisition of the same fluorescence response matrix with the same scan grid, exposure per position and detection region. Here we also give an idealized acquisition-count comparison with laser-scanning microscopy (LSM) and confocal laser-scanning microscopy (CLSM). Their counts refer to point measurements, whereas the single-focus and multifocal counts refer to camera frames, so the counts do not translate into equal detector readout times. We assume a square grid, neglect edge effects and take the excitation and emission wavelengths to be approximately equal.

For this comparison, all acquisition methods use the same scan step:

$$\Delta x_{\mathrm{LSM/CLSM}} \approx \Delta x_{\mathrm{Multi}} \approx \frac{\lambda_{\mathrm{ex}}}{4\mathrm{NA}}. \tag{S8}$$

The experimental scan step must be chosen for the relevant optical bandwidth, detection geometry and confocal pinhole size[1,2]. At a matched scan step, the LSM/CLSM point count is the imaged area divided by the squared scan step, and single-focus acquisition requires the same number of camera frames.

In multifocal acquisition, $M$ foci measure $M$ scan positions per exposure. Excluding calibration frames, the ideal number of frameshifts is

$$N_{\mathrm{Multi}} \approx \frac{A}{M\Delta x_{\mathrm{Multi}}{}^{2}} \approx \frac{16A\mathrm{NA}^{2}}{M\lambda_{\mathrm{ex}}{}^{2}}. \tag{S9}$$

For a single-focus reference B at scan step $\Delta x_{\mathrm{B}}$, the general acquisition-count ratio is

$$G_{N,\mathrm{B}} \equiv \frac{N_{\mathrm{B}}}{N_{\mathrm{Multi}}} \approx M\left(\frac{\Delta x_{\mathrm{Multi}}}{\Delta x_{\mathrm{B}}}\right)^{2}. \tag{S10}$$

At a matched scan step, this ratio reduces to the $M$-fold parallelization gain[3,4]; no additional gain is attributed to a denser scan grid in the single-focus reference. The reduction in elapsed time also depends on exposure, pattern switching, camera readout, data transfer and spacing calibration. Increasing the lattice spacing reduces $M$ and therefore increases the number of frameshifts; any further effect on image SNR depends on how the contributing measurements are combined (see Noise accumulation in frame-summed measurements below).

**Reference comparison and local OTF estimation for the strongly aberrated dataset**

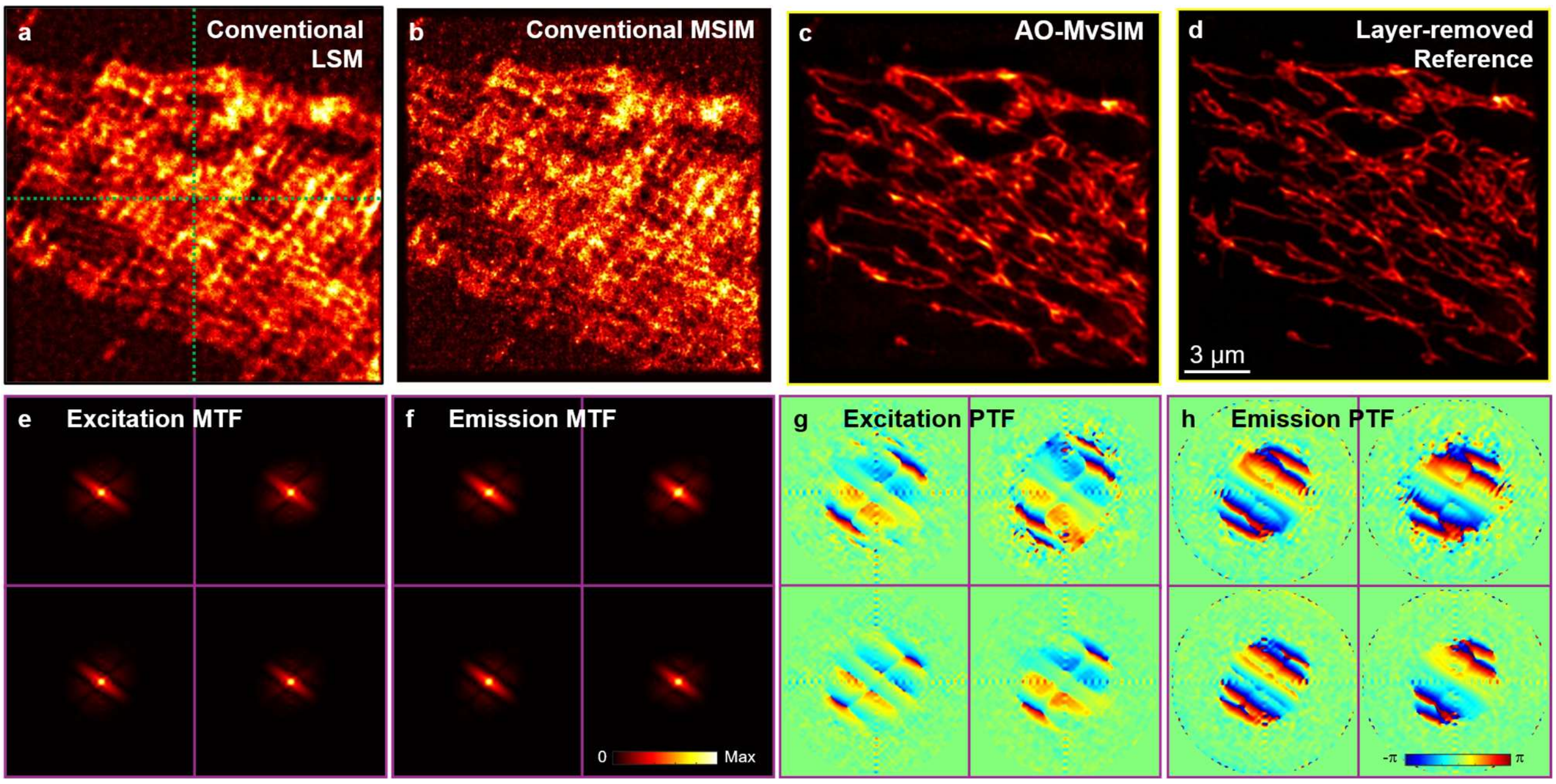

**Fig. S5 | Reference comparison and local OTF estimates for the strongly aberrated BPAE-cell dataset. a–d**, Conventional LSM (**a**), conventional MSIM (**b**), AO-MvSIM (**c**) and the independently acquired layer-removed reference (**d**) for the mitochondrial region examined in Fig. 3. Green dotted lines in **a** divide the field into the four regions used for local OTF estimation. **e**,**f**, Recovered excitation and emission MTFs. **g**,**h**, Corresponding excitation and emission PTFs. Each set of transfer functions is arranged in the same 2 × 2 order as the image regions. The MTFs show anisotropic frequency transmission, and the PTFs show the distinct phase structures estimated for the two paths. MTF magnitude is shown from zero to its displayed maximum, and phase from −π to π. Scale bar, 3 µm in **d**.

Figure S5 compares conventional LSM, conventional MSIM and AO-MvSIM with the layer-removed reference. AO-MvSIM preserves the continuous mitochondrial network, whereas the conventional reconstructions remain diffuse. The four local excitation and emission OTF estimates are the responses used for path-resolved correction, and the layer-removed image serves as the structural reference for judging recovery of the network.

**Characterization and correction of excitation and emission aberrations in BPAE cells**

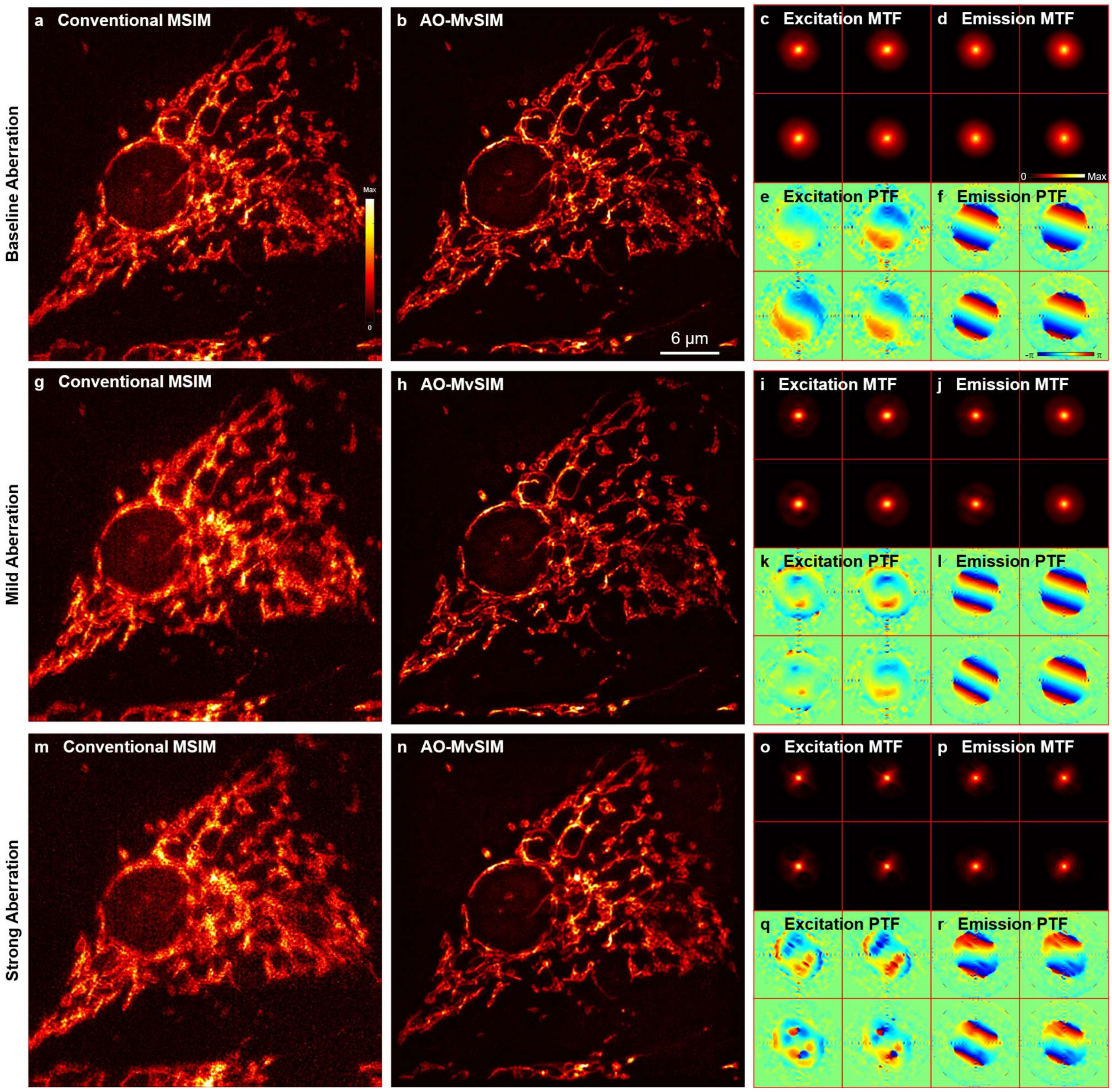

**Fig. S6 | Local excitation and emission transfer functions under different aberration conditions.** The same BPAE-cell region was imaged under the baseline condition (**a**–**f**), with an artificial aberrating layer adding mild distortion (**g**–**l**) and after translating that layer to add stronger distortion (**m**–**r**). **a**,**b**, Conventional MSIM and AO-MvSIM images under the baseline condition. **c**,**d**, Corresponding excitation and emission MTFs. **e**,**f**, Corresponding excitation and emission PTFs. **g**,**h**, Conventional MSIM and AO-MvSIM images with mild additional aberration. **i**,**j**, Corresponding excitation and emission MTFs; **k**,**l**, excitation and emission PTFs. **m**,**n**, Conventional MSIM and AO-MvSIM images with stronger aberration. **o**,**p**, Corresponding excitation and emission MTFs; **q**,**r**, excitation and emission PTFs. Each transfer-function panel contains the estimates for four local patches in a 2 × 2 arrangement. Images show the mitochondrial channel. MTFs display the OTF magnitude and PTFs the OTF phase from −π to π. Scale bar, 6 μm in **b**.

The baseline images in Fig. S6a,b show the same BPAE-cell region reconstructed by conventional MSIM and AO-MvSIM. The excitation and emission MTFs are compact and nearly radially symmetric, whereas the PTFs differ between the two paths (Fig. S6c–f). Adding the aberrating layer changes the recovered magnitude and phase distributions (Fig. S6i–l), and translating it to the stronger-aberration condition produces more anisotropic frequency transmission and more complex phase structure (Fig. S6o–r). AO-MvSIM separates neighboring mitochondria more clearly under all three conditions (Fig. S6b,h,n).

For the strongly aberrated dataset of Fig. 3, the four local excitation and emission MTFs and PTFs are shown in Fig. S5e–h. The excitation and emission responses differ and show anisotropic modulation and structured phase. Each local factorization leaves an undetermined linear phase common to the two OTFs, so patch-to-patch differences in linear phase alone cannot establish spatial variation of the physical aberration. Recovered detail remains limited to the spatial frequencies supported by the measured fluorescence.

**Dendritic spine detectability in different reconstructions of the same raw data**

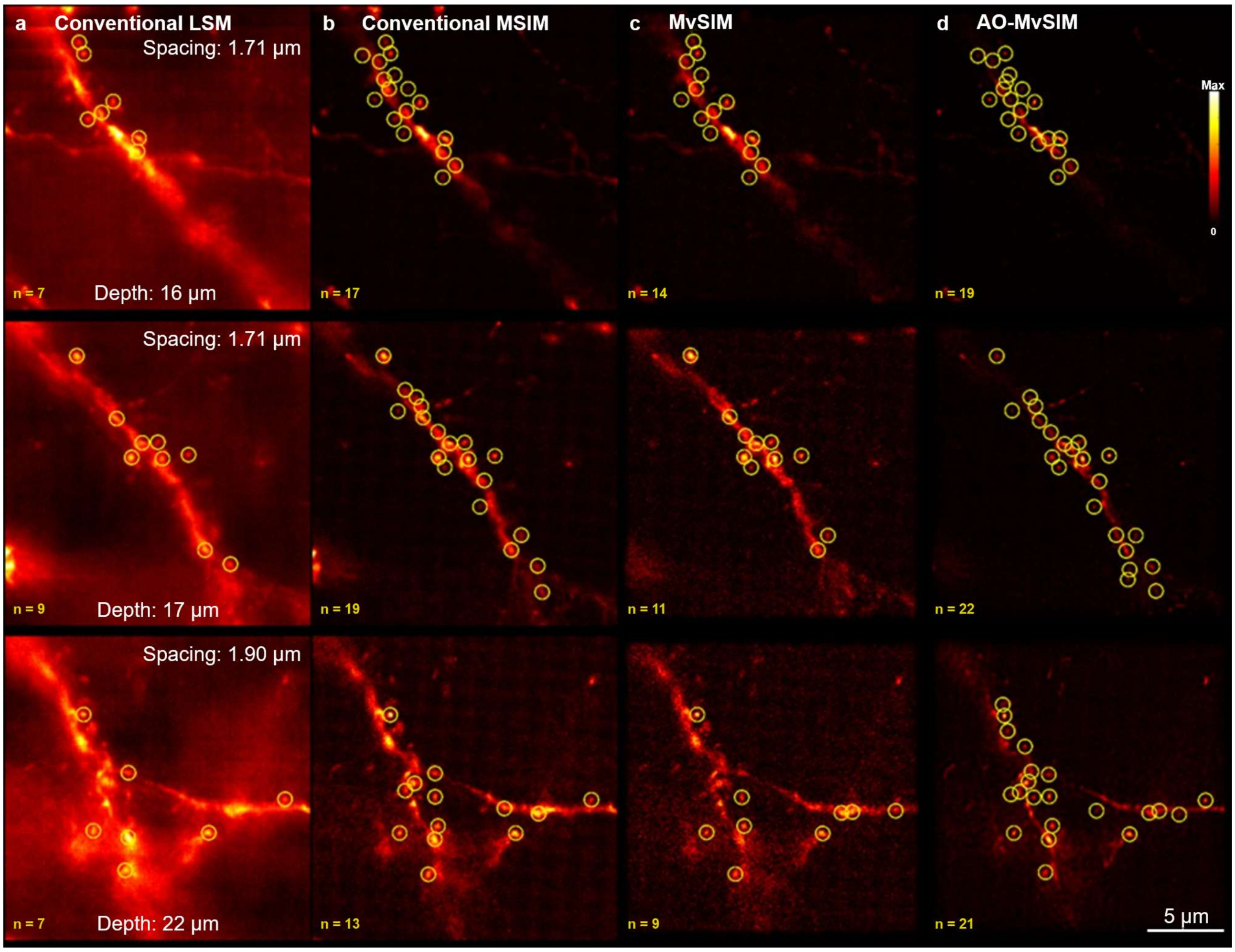

**Fig. S7 | Dendritic spine detectability across reconstructions of the same raw datasets.** Columns **a**–**d** show conventional LSM, conventional MSIM, MvSIM and AO-MvSIM reconstructions, respectively. Rows correspond to depths of 16, 17 and 22 µm, with lattice spacings of 1.71, 1.71 and 1.90 µm. Yellow circles mark the identified protrusions, and *n* gives the count in each panel: 7/17/14/19 at 16 µm, 9/19/11/22 at 17 µm and 7/13/9/21 at 22 µm across **a**–**d**. Detection parameters and dendritic centerlines were shared within each field (Table S2 and the procedure below). Centerlines were traced on the AO-MvSIM images, so the counts describe relative detectability rather than accuracy against independent ground truth. Scale bar, 5 µm in the bottom-right panel.

The three fields contain 23, 49, 34 and 62 panel-specific detections in the conventional LSM, conventional MSIM, MvSIM and AO-MvSIM images, respectively; relative to AO-MvSIM, the three comparison methods detect about 37%, 79% and 55% (Table S2). These fractions measure relative detectability under the stated procedure, not sensitivity against ground truth. Spine-neck profiles are reported separately in Fig. 4.

**Table S2 | Dendritic spine counts and traced lengths**

| Imaging depth | Lattice spacing (μm) | Traced length (μm) | Conventional LSM (a) | Conventional MSIM (b) | MvSIM (c) | AO-MvSIM (d) |
|---|---|---|---|---|---|---|
| **16 μm** | 1.71 | 11.3 | 7 | 17 | 14 | 19 |
| **17 μm** | 1.71 | 20.0 | 9 | 19 | 11 | 22 |
| **22 μm** | 1.90 | 28.4 | 7 | 13 | 9 | 21 |
| **Total spines** | **—** | **59.7** | **23** | **49** | **34** | **62** |
| Detected relative to corrected reconstruction (%) | — | — | 37 | 79 | 55 | 100 |

For each field of view, the dendritic shaft centerline was traced once on the AO-MvSIM panel and applied to all four panels of that row; where more than one dendrite was in focus (bottom row), every in-focus shaft was traced. Candidate spine heads were found with a difference-of-Gaussians filter ($\sigma_1 = 0.067$ μm, $\sigma_2 = 0.19$ μm) applied to the intensity map. A local maximum was kept as a candidate when it satisfied all three criteria below, with the same parameters for all twelve panels:

(I) Its filter response exceeded the background level by at least 20 × the median absolute deviation of the response, with background estimated from all pixels more than 4 μm from any traced centerline in the same panel;

(II) It lay between 0.35 and 2.1 μm from the nearest traced centerline, i.e., off the shaft but within the specified search range;

(III) It was separated by at least 0.38 μm from every other accepted maximum.

Every candidate was then inspected at high magnification and rejected if it was a diffuse feature in the out-of-focus background with no visible connection to a traced shaft. Four of 172 candidates were rejected, leaving the 168 panel-specific detections marked in Fig. S7. Because the centerlines were traced on the AO-MvSIM images, the distance criterion is defined relative to that reconstruction.

**Optical configuration for multifocal fluorescence response matrix acquisition**

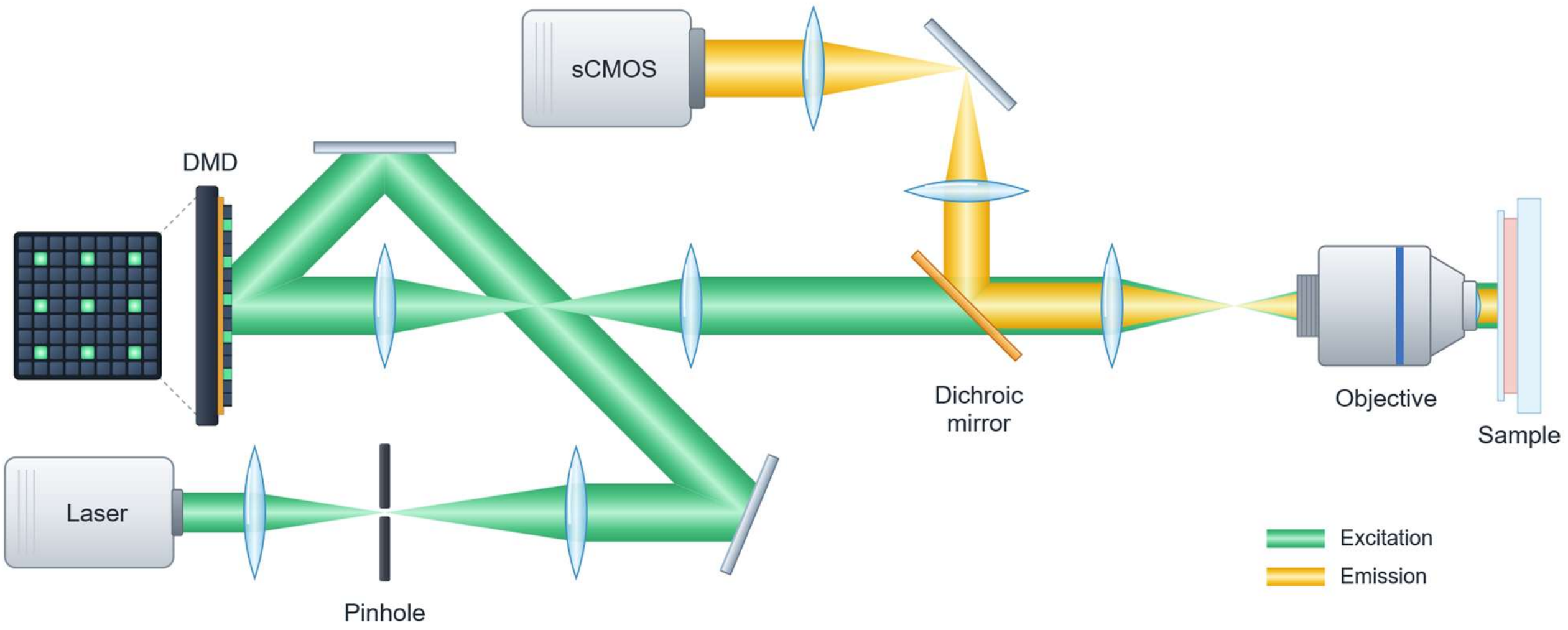

**Fig. S8 | Optical configuration for DMD-based multifocal fluorescence response matrix acquisition.** The laser beam is spatially filtered by a pinhole and relayed to the DMD. The programmed micromirror pattern generates the multifocal lattice, which is projected onto the specimen through the relay optics and the objective. Fluorescence returns through the same objective, is separated from the excitation by the dichroic mirror and is imaged onto the sCMOS camera. Green and yellow denote the excitation and emission paths, respectively. The inset shows the array of active DMD elements.

**Fluorescence excitation:** Two continuous-wave lasers were used: a 488-nm Cobolt 06-MLD for EGFP, Alexa Fluor 488-labeled specimens and yellow-green fluorescent beads, and a 561-nm OBIS LS 561-150 for specimens labeled with MitoTracker Red.

**Excitation beam spatial filtering:** A spatial filter consisting of a pinhole at the common focal plane of two lenses removed unwanted spatial components of the excitation beam before the DMD.

**Beam delivery and DMD operation:** Two mirrors directed the beam to the DMD (DLI4120VIS-951080P, Digital Light Innovations, with a Texas Instruments DMD). Binary patterns streamed from the host computer as 8-bit image buffers set the on/off state of each micromirror; on-state mirrors directed light into the microscope and off-state mirrors directed it out of the imaging path.

**Separation of excitation and emission:** A multiband dichroic beamsplitter transmitted the excitation light toward the objective and reflected the fluorescence into the detection path.

**Multifocal lattice scanning:** The DMD shifted the multifocal lattice over a two-dimensional raster grid, so the specimen was scanned by updating the DMD pattern without galvanometric mirrors[5,6].

**Objective lenses:** Fluorescence was excited and collected through a water-dipping objective (N60X-NIR, Nikon; 60×, NA 1.0) or a silicone oil-immersion objective (CFI SR HP Plan Apochromat Lambda S 100XC Sil, Nikon; 100×, NA 1.35).

**Fluorescence filtering and camera detection:** Fluorescence collected by the objective was reflected by the dichroic beamsplitter into the detection path, passed through a multiband emission filter (ZET488/561m-TRF, Chroma) that rejected residual excitation light, and was imaged onto a scientific CMOS (sCMOS) camera (pco.edge 4.2, PCO AG). One camera frame was recorded for each DMD pattern, that is, for each frameshift.

**Fluorescence response matrix formulation**

The reconstructed responses were arranged into the spatial-domain fluorescence response matrix

$$\mathbf{F}^{(\mathrm{r})} = [f^{(\rho)}(\mathbf{r}_{\mathrm{d},m}, \mathbf{r}_{\mathrm{i},n})]_{mn}. \tag{S11}$$

Here $m = 1, \ldots, N_{\mathrm{d}}$ and $n = 1, \ldots, N_{\mathrm{i}}$ index the detection and illumination positions; these matrix indices are distinct from the lattice parameters of the reconstruction protocol. Each response is expressed on the common detector grid, with positions outside its ROD set to zero. Explicitly,

$$\mathbf{F}^{(\mathrm{r})}=\begin{bmatrix} f^{(\rho)}(\mathbf{r}_{\mathrm{d},1}, \mathbf{r}_{\mathrm{i},1}) & f^{(\rho)}(\mathbf{r}_{\mathrm{d},1}, \mathbf{r}_{\mathrm{i},2}) & \cdots & f^{(\rho)}(\mathbf{r}_{\mathrm{d},1}, \mathbf{r}_{\mathrm{i},N_{\mathrm{i}}}) \\ f^{(\rho)}(\mathbf{r}_{\mathrm{d},2}, \mathbf{r}_{\mathrm{i},1}) & f^{(\rho)}(\mathbf{r}_{\mathrm{d},2}, \mathbf{r}_{\mathrm{i},2}) & \cdots & f^{(\rho)}(\mathbf{r}_{\mathrm{d},2}, \mathbf{r}_{\mathrm{i},N_{\mathrm{i}}}) \\ \vdots & \vdots & \ddots & \vdots \\ f^{(\rho)}(\mathbf{r}_{\mathrm{d},N_{\mathrm{d}}}, \mathbf{r}_{\mathrm{i},1}) & f^{(\rho)}(\mathbf{r}_{\mathrm{d},N_{\mathrm{d}}}, \mathbf{r}_{\mathrm{i},2}) & \cdots & f^{(\rho)}(\mathbf{r}_{\mathrm{d},N_{\mathrm{d}}}, \mathbf{r}_{\mathrm{i},N_{\mathrm{i}}}) \end{bmatrix}. \tag{S12}$$

Each column contains the detector-resolved response for one illumination position,

$$\mathbf{f}_n=\begin{bmatrix} f^{(\rho)}(\mathbf{r}_{\mathrm{d},1}, \mathbf{r}_{\mathrm{i},n}) \\ f^{(\rho)}(\mathbf{r}_{\mathrm{d},2}, \mathbf{r}_{\mathrm{i},n}) \\ \vdots \\ f^{(\rho)}(\mathbf{r}_{\mathrm{d},N_{\mathrm{d}}}, \mathbf{r}_{\mathrm{i},n}) \end{bmatrix},$$

$$\mathbf{F}^{(\mathrm{r})}=[\mathbf{f}_1 \quad \mathbf{f}_2 \quad \cdots \quad \mathbf{f}_{N_{\mathrm{i}}}]. \tag{S13}$$

Fourier transformation over both two-dimensional coordinate sets yields the spectral response matrix $\mathbf{F}^{(\mathrm{k})}$, whose scalar elements are denoted $F^{(\mathrm{k})}(\mathbf{k}_{\mathrm{d}}, \mathbf{k}_{\mathrm{i}})$. The factorized model below assumes that residual crosstalk and truncation by the finite ROD have negligible effects on the retained response.

**Patch-wise dual deconvolution**

Each locally shift-invariant patch was processed independently with the fluorescence response model[7]. Let $O(\mathbf{r})$ denote the fluorophore distribution, with the PSFs defined by their displacement from the illumination and detection positions. The local forward model is

$$f(\mathbf{r}_{\mathrm{d}}, \mathbf{r}_{\mathrm{i}}) \approx \int h_{\mathrm{em}}(\mathbf{r} - \mathbf{r}_{\mathrm{d}}) O(\mathbf{r}) h_{\mathrm{ex}}(\mathbf{r} - \mathbf{r}_{\mathrm{i}}) \mathrm{d}^2\mathbf{r}. \tag{S14}$$

With this displacement convention, the excitation and emission OTFs are the positive-exponent Fourier transforms of $h_{\mathrm{ex}}$ and $h_{\mathrm{em}}$, and transforming over both detection and illumination coordinates gives

$$F^{(\mathrm{k})}(\mathbf{k}_{\mathrm{d}},\mathbf{k}_{\mathrm{i}}) \approx H_{\mathrm{em}}(\mathbf{k}_{\mathrm{d}})\, \Gamma(\mathbf{k}_{\mathrm{d}} + \mathbf{k}_{\mathrm{i}})\, H_{\mathrm{ex}}(\mathbf{k}_{\mathrm{i}}). \tag{S15}$$

Here $\Gamma$ is the object spectrum. We write $F_{\mathrm{d,i}} = F^{(\mathrm{k})}(\mathbf{k}_{\mathrm{d}}, \mathbf{k}_{\mathrm{i}})$ and $H_{\mathrm{d,i}} = H_{\mathrm{em}}(\mathbf{k}_{\mathrm{d}})\, H_{\mathrm{ex}}(\mathbf{k}_{\mathrm{i}})$. Starting from unit OTF estimates, alternating Wiener-type updates enforce a common object-spectrum value for all coordinate pairs with $\Delta\mathbf{k} = \mathbf{k}_{\mathrm{d}} + \mathbf{k}_{\mathrm{i}}$. The object update is

$$\Gamma_{\mathrm{syn}}(\Delta\mathbf{k}) \leftarrow \frac{H_{\mathrm{eff}}(\Delta\mathbf{k})}{{H_{\mathrm{eff}}}^2(\Delta\mathbf{k}) + {\varepsilon_{\mathrm{eff}}}^2} \sum_{\mathbf{k}_{\mathrm{i}}} \frac{\overline{H_{\mathrm{d,i}}}\, F_{\mathrm{d,i}}}{\left|H_{\mathrm{d,i}}\right|^2 + {\varepsilon_{\Gamma}}^2}, \tag{S16}$$

with the effective spectral weighting

$$H_{\mathrm{eff}}(\Delta\mathbf{k}) = \sum_{\mathbf{k}_{\mathrm{i}}} \frac{|H_{\mathrm{d,i}}|^2}{|H_{\mathrm{d,i}}|^2 + {\varepsilon_{\Gamma}}^2}. \tag{S17}$$

In these sums $\mathbf{k}_{\mathrm{d}} = \Delta\mathbf{k} - \mathbf{k}_{\mathrm{i}}$, so each sum combines frequency pairs with the same object frequency. After setting $\Gamma_{\mathrm{d,i}} \leftarrow \Gamma_{\mathrm{syn}}(\mathbf{k}_{\mathrm{d}} + \mathbf{k}_{\mathrm{i}})$, the emission OTF is updated as

$$H_{\mathrm{em}}(\mathbf{k}_{\mathrm{d}}) \leftarrow \frac{S_{\mathrm{em}}(\mathbf{k}_{\mathrm{d}})}{{S_{\mathrm{em}}}^2(\mathbf{k}_{\mathrm{d}}) + {\varepsilon_{\mathrm{H}}}^2} \sum_{\mathbf{k}_{\mathrm{i}} \neq 0} F_{\mathrm{d,i}}\, \overline{\Gamma_{\mathrm{d,i}}\, H_{\mathrm{ex}}(\mathbf{k}_{\mathrm{i}})}, \tag{S18}$$

where $S_{\mathrm{em}}(\mathbf{k}_{\mathrm{d}}) = \sum_{\mathbf{k}_{\mathrm{i}} \neq 0} |\Gamma_{\mathrm{d,i}}\, H_{\mathrm{ex}}(\mathbf{k}_{\mathrm{i}})|^2$. The excitation update has the corresponding form

$$H_{\mathrm{ex}}(\mathbf{k}_{\mathrm{i}}) \leftarrow \frac{S_{\mathrm{ex}}(\mathbf{k}_{\mathrm{i}})}{{S_{\mathrm{ex}}}^2(\mathbf{k}_{\mathrm{i}}) + {\varepsilon_{\mathrm{H}}}^2} \sum_{\mathbf{k}_{\mathrm{d}} \neq 0} F_{\mathrm{d,i}}\, \overline{\Gamma_{\mathrm{d,i}}\, H_{\mathrm{em}}(\mathbf{k}_{\mathrm{d}})}. \tag{S19}$$

Here $S_{\mathrm{ex}}(\mathbf{k}_{\mathrm{i}}) = \sum_{\mathbf{k}_{\mathrm{d}} \neq 0} |\Gamma_{\mathrm{d,i}} H_{\mathrm{em}}(\mathbf{k}_{\mathrm{d}})|^2$, and overbars denote complex conjugation. The regularization parameters $\varepsilon_{\Gamma}$, $\varepsilon_{\mathrm{eff}}$ and $\varepsilon_{\mathrm{H}}$ were chosen empirically according to the SNR. The object spectrum and the two OTF estimates were updated alternately; five iterations were used for the kidney reconstruction timed in the main text. Inverse Fourier transformation of $\Gamma_{\mathrm{syn}}$ gives the AO-MvSIM image, and the OTF magnitudes and phases give the MTFs and PTFs. The datasets of Figs. 3 and 5 were processed in locally shift-invariant patches.

**Acquisition and reconstruction parameters for Figs. 2, 3 and 5**

**Table S3 | Acquisition and reconstruction parameters for Figs. 2, 3 and 5**

| **Parameter** | **Symbol** | **Fig. 2** | **Figs. 3 and 5** |
|---|---|---|---|
| Dark spacing between patches | $m$ | 28 pixels | 48 pixels |
| Lattice period | $p = m + 2$ | 30 pixels | 50 pixels |
| Foci per axis | $n$ | 34 | 20 |
| Foci per frameshift | $n^2$ | 1,156 | 400 |
| Number of frameshifts | $N_{\text{frame}} = p^2$ | 900 | 2,500 |
| Scan-grid width | $N = np$ | 1,020 pixels | 1,000 pixels |
| Effective illumination positions | $N^2$ | 1,040,400 | 1,000,000 |
| ROD width | $\rho$ | 30 pixels | 50 pixels |
| ROD margin | $\delta_\rho = (\rho - 2)/2$ | 14 pixels | 24 pixels |
| Local detector samples per ROD | $\rho^2$ | 900 | 2,500 |
| Cropped response stack | $\mathbf{R}^{(\rho)}$ | 30 × 30 × 1,040,400 | 50 × 50 × 1,000,000 |

Pixel values refer to the registered DMD grid. Counts assume full $n \times n$ lattices and $p^2$ shifts ($n^2p^2 = N^2$); positions whose RODs extend beyond the camera field are masked or excluded. The listed stack dimensions refer to the cropped responses before reindexing to absolute coordinates.

**Noise accumulation in frame-summed measurements**

Adding frames lowers the SNR when they contribute background and read noise without additional signal. Consider a sum with total signal $S$ (photoelectrons) from $n$ frames at one detector pixel, with background $b$ per frame and read-noise variance $\sigma_{\mathrm{r}}^2$ per frame. For Poisson statistics and independent read noise[8], after subtracting the mean background,

$$\mathrm{SNR} = \frac{S}{\sqrt{S + n(b + \sigma_{\mathrm{r}}^2)}}. \tag{S20}$$

Subtraction removes the mean background but not its shot noise; a residual error $\Delta b$ per frame would add a bias of $n\Delta b$, and uncertainty in the background estimate would add further noise (not included here). The count $n$ includes only the frames that enter the sum.

When $n(b + \sigma_{\mathrm{r}}^2) \gg S$, the SNR falls as $n^{-1/2}$ for fixed $S$, $b$ and $\sigma_{\mathrm{r}}^2$. In our reconstruction, each illumination position uses only the corresponding frameshift and ROD, so more frameshifts do not automatically increase $n$; how the lattice spacing affects the final SNR depends on frame selection, exposure, dose and background. The expression applies to a linear sum before clipping or deconvolution.

**Synchronization of DMD patterns and camera acquisition**

**Hardware and software configuration:**

The hardware and software were configured as follows:

1. Launch Micro-Manager 2 and connect the sCMOS camera (pco.edge 4.2, rolling-shutter mode) as the active imaging device.

2. Add the DMD to the Micro-Manager 2 device configuration.

3. Load the multifocal illumination pattern stack into MATLAB memory.

4. Configure the camera in Micro-Manager 2 for external transistor–transistor logic (TTL) hardware triggering.

After configuration, the MATLAB script loaded one pattern onto the DMD, sent a TTL pulse to trigger the camera and waited for exposure and readout to finish before advancing to the next pattern. The sequence is listed below.

**Pattern update and camera triggering sequence:**

| Step | Operation |
|---|---|
| 1 | Initialization: Sequential acquisition starts |
| 2 | Create a C-compatible pointer for dynamic-link library (DLL) communication |
| 3 | Load the DMD row buffer via the ALP device library |
| 4 | Reset the DMD to the specified address |
| 5 | Send a TTL HIGH pulse via the data acquisition (DAQ) output channel |
| 6 | Trigger camera frame acquisition |
| 7 | Hold TTL HIGH for the trigger pulse width |
| 8 | Send TTL LOW to end the trigger pulse |
| 9 | Wait for camera exposure and readout to complete |
| 10 | Advance to the next pattern |

The acquisition duty cycle depends on the pattern-transfer time, the trigger sequence, camera exposure and readout.